\documentclass{ws-ijgmmp}

\usepackage{xcolor}
\usepackage{rotating}
\usepackage{tikz}
\usetikzlibrary{arrows.meta, positioning}

\usepackage[verbose,hypertexnames=false]{hyperref}
\hypersetup{colorlinks=false,allbordercolors=blue,pdfborderstyle={/S/U/W 1}}
\usepackage[numbers,sort&compress]{natbib}
\begin{document}

\markboth{Samstuti Chanda \& Ranjan Sharma }
{Two-Parameter Deformation of Embedding Class-I Compact Stars in Linear $f(Q)$ Gravity}

%
\catchline{}{}{}{}{}
%

\title{Two Parameter Deformation of Embedding Class-I Compact Stars in Linear $f(Q)$ Gravity}

\author{Samstuti Chanda\footnote{Permanent affiliation: Department of Physics, Islampur College, Islampur, 733202, West Bengal, India.}}

\address{IUCAA Centre for Astronomy Research and Development, Department of Physics, Cooch Behar Panchanan Barma University, Cooch Behar, 736101, West Bengal, India\\
\email{schanda93.dta@gmail.com}}

\author{Ranjan Sharma}
\address{IUCAA Centre for Astronomy Research and Development, Department of Physics, Cooch Behar Panchanan Barma University, Cooch Behar, 736101, West Bengal, India\\
\email{rsharma@associates.iucaa.in}}

\maketitle

\begin{history}
\received{(Day Month Year)}
\revised{(Day Month Year)}
\accepted{(Day Month Year)}
\published{(Day Month Year)}
\end{history}

\begin{abstract}

Recent multi-messenger observations, including gravitational wave detections of compact objects in the neutron star-black hole mass-gap region and precise measurements of high-mass pulsars, motivate mechanisms that can enlarge the stellar mass window without arbitrarily stiffening the equation of state (EOS) toward the causal limit. In linear $f(Q)$ gravity of the form $f(Q)=\beta_1 Q+\beta_2$, the theory is dynamically equivalent to General Relativity (GR) at the geometric level and modifies stellar structure solely through a uniform rescaling of the matter sector governed by $\beta_1$. Consequently, linear $f(Q)$ alone does not introduce new geometric families of stellar solutions or alter classical compactness bounds. To overcome this structural limitation, we incorporate gravitational decoupling within an embedding class-I (Karmarkar) Vaidya-Tikekar configuration in linear $f(Q)$ gravity. While similar VT-based decoupling constructions exist in GR, the present framework introduces a controlled two-parameter deformation characterized by $(\epsilon,~\beta_1)$: the decoupling parameter $\epsilon$ governs geometric deformation and EOS stiffness, whereas $\beta_1$ independently rescales the matter sector without altering the metric structure. This separation permits a direct comparison between GR and linear $f(Q)$ gravity at fixed geometric deformation, thereby isolating pure coupling-driven mass enhancement. We determine the admissible parameter domain based on regularity, matching, causality and compactness requirements, and derive an analytic compactness bound for the decoupled embedding class-I configuration. We show that, at fixed $\epsilon$ (i.e., fixed effective stiffness), linear $f(Q)$ systematically shifts the mass-radius sequence relative to GR through matter-sector rescaling alone. The combined action of $\epsilon$ and $\beta_1$ enlarges the accessible stellar mass window while preserving physical acceptability, allowing configurations compatible with recent high-mass pulsars and mass-gap candidates without exceeding causal limits.

\end{abstract}



\section{Introduction}\label{sec1}

The construction of a realistic compact star model is governed not only by the microphysics of dense matter but also, more fundamentally, by the geometric assumptions imposed on the interior spacetime. Embedding class-I geometry and the associated Karmarkar condition~\cite{karmarkar} provides a powerful framework for reducing arbitrariness in relativistic stellar models. The Karmarkar condition imposes a geometric restriction on the metric potentials of a static, spherically symmetric spacetime, requiring that they be compatible with an embedding in a five-dimensional flat manifold. In differential geometric terms, it provides a condition for a class-II spacetime to be reduced to embedding class~I, effectively constraining the Riemann tensor components so that only one metric function remains independent. This reduction greatly simplifies the solution-generating technique and is useful, particularly for constructing relativistic stellar interiors.  A wide range of solutions satisfying the Karmarkar condition have been developed, including anisotropic~\cite{bhar0,bhar1,bhar2,bhar3,maur1,maur2,tello,ratan,Singh2016,Singh2020}, charged~\cite{bhar1,makalo2023,Estevez2023,GomezLeyton2021} and wormhole solutions~\cite{sutar2023,fayyaz2020,malik2024} with further extensions to modified gravity frameworks such as energy-momentum squared gravity~\cite{Sharif2023} and Rastall gravity~\cite{Zubair2021}. In a recent paper, Chanda \textit{et al}~\cite{Chanda2026} showed that, when embedding class-I geometry is treated as a rigid curvature condition, especially when combined with additional geometric restrictions, the resulting spacetime becomes highly restrictive, admitting only idealized isotropic configurations with vanishing anisotropy and complexity. The analysis demonstrated that severe limitations might arise purely at the geometric level, even before specifying any equation of state (EOS) or matter composition.

It is noteworthy that recent multi-messenger observations demand greater structural flexibility in compact star models. Measurements of heavy pulsars with masses close to or exceeding $2M_\odot$, together with emerging evidence for objects in the neutron star-black hole mass-gap region, challenge models constructed within overly constrained geometric frameworks. These observations motivate the search for mechanisms capable of extending the stellar mass range while maintaining physical consistency and avoiding unrealistic stiffening of the equation of state.

Within modified gravity, symmetric teleparallel gravity and its $f(Q)$ extension provide a particularly transparent setting for such investigations. The symmetric teleparallel formulation of gravity describes gravitation through the non-metricity scalar $Q$ rather than curvature or torsion. This framework offers a comparatively simple geometric structure compared to curvature-based extensions (e.g., $f(R)$ gravity) or torsion-based alternatives (e.g., $f(T)$ \cite{Jimenez2019}). In the coincident gauge, the affine connection can be set to zero, forcing both curvature and torsion to vanish and leaving all gravitational information encoded in the non-metricity. The resulting field equations remain second-order in the metric, which avoids the higher-order instabilities often encountered in modified curvature theories. Extending this construction to $f(Q)$ gravity by promoting the Lagrangian to a general function of $Q$ preserves this second-order character and provides a flexible theoretical framework \cite{Lazkoz2019}. Cosmological studies reinforce the viability of $f(Q)$, showing that suitable choices of the function reproduce late-time cosmic acceleration, mimic diverse effective dark-energy behaviours, and remain compatible with both background and perturbative observational constraints \cite{Lazkoz2019,Ayuso2021,Barros2020}. These cosmological motivations establish a natural basis for investigating stellar systems in the same geometric setting.

In stellar modelling, the theoretical structure of $f(Q)$ gravity was clarified by Wang \textit{et al}~\cite{Wang2022}, who analyzed static spherically symmetric stellar configurations and showed that Schwarzschild or Schwarzschild-(anti-)de Sitter exterior solutions arise only when the gravitational Lagrangian is strictly linear in the non-metricity scalar. This matching requirement immediately constrains the choice of the function $f(Q)$ in the form $f(Q)=\beta_1Q+\beta_2$, where $\beta_1$ and $\beta_2$ are two constant parameters, as the only viable option for compact star modelling.  Subsequent studies assuming isotropic~\cite{maurya2024fq} and anisotropic fluid spheres ~\cite{adeel2023fq,maurya2024cqg,maurya2024jcap,errehymy2024fq,paul2024fq,kumar2025fq,Sharma2024,Awais2025,bhar2024cjph,Rani2024}, dark energy stars (DES)~\cite{bhar2023dark,bhar2023fq}, strange quark stars~\cite{bhar2023dark}, hybrid quark-hadron~\cite{bhar2023hybrid} configurations and charged stellar models~\cite{maurya2024prop,kumar2024fq,maurya2024epjcfq,rani2024fq,shahzad2025}, show how non-metricity modifies structural behaviour and reshapes pressure anisotropy and stability conditions while preserving physical viability. As observational constraints tightened, several of these works constructed mass-radius relations and maximum mass sequences within linear $f(Q)$, highlighting systematic departures from General Relativistic (GR) predictions. However, recent studies by De and Loo~\cite{De2023} showed that only linear $f(Q)$ models generically preserve covariant conservation of the energy-momentum tensor. In contrast, non-linear forms violate the conservation principle unless $Q$ is assumed to be a constant, effectively reducing the theory to GR with a cosmological constant. Heisenberg and Pastor-Marcos~\cite{Heisenberg2025} demonstrated that non-linear extensions tend to recover GR-like behaviour for compact objects under standard assumptions, unless the affine connection is treated dynamically. These results motivate adopting linear $f(Q)$ gravity as a theoretically consistent baseline. Nevertheless, since linear $f(Q)$ gravity is geometrically equivalent to GR, such constructions remain structurally restrictive. In other words, a linear $f(Q)$ modification introduces no independent geometric deformation beyond matter normalization. Consequently, linear $f(Q)$ gravity, by itself, lacks the capacity to generate new geometric families of stellar configurations or to modify classical compactness bounds, even though it may shift the overall mass scale.

The above observations reveal a structural limitation for embedding class-I spacetime in linear $f(Q)$ gravity. Class-I geometry restricts metric freedom, while linear $f(Q)$ gravity preserves the GR geometric sector and modifies only matter normalization. The combined framework, therefore, remains effectively a single-parameter framework, governed by the coupling constant $\beta_1$. To enlarge the solution space beyond pure matter rescaling, an additional independent geometric deformation mechanism is required. Gravitational decoupling (GD), realized through minimal geometric deformation (MGD) and complete geometric deformation (CGD), provides precisely such a mechanism. Originally developed for spherically symmetric brane-world systems, the MGD method was extended to allow deformation of both the time and radial metric components, yielding a modified Schwarzschild exterior and a new exact solution for stellar objects in the extra-dimensional scenario~\cite{casadio2015}. Ovalle demonstrated that one can start from an isotropic seed solution in GR and decouple an additional gravitational source via a controlled deformation of the metric potentials, yielding anisotropic matter distributions while ensuring regularity and physical viability~\cite{ovalle2017}. In subsequent studies, many investigators have applied this approach to $(2+1)$ dimensional~\cite{contreras2018}, stellar interiors including neutron stars~\cite{ovalle2019,torres2019,abellan2020}, higher-order gravity theories~\cite{sultana2021}, ultra-compact charged and mass-gap configurations~\cite{maurya2020epjcnon,maurya2025decoupling,maurya2020mgd,maurya2020rastall,maurya2024apj,maurya2025vc,maurya2024jheap,mushtaq2025ctp,pradhan2024fqt,maurya2023df,maurya2023egb,maurya2023buchdahl,maurya2023vcf,albusaidi2023,maurya2023mgd0,jasim2023bd,Dayanandan2023CJP,Habsi2023EPJC286,maurya2023mnras,alhadhrami2023,maurya2022fs,maurya2022classI,maurya2022cpc,maurya2022ps,maurya2022egb00,maurya2022complexity}. Although GD conventionally assumes an isotropic seed \cite{ovalle2017}, several instances employing anisotropic seeds are also known. In $f(Q)$ gravity, both isotropic~\cite{maurya2023cf,maurya2024decouple} and anisotropic~\cite{gd_darkmatter2024,gd_strangestars2023,maurya2022strange} seeds have been supplemented via GD.

 Very recently, the Karmarkar condition has been applied to $f(Q)$ gravity itself, obtaining anisotropic stars constrained by observed masses and radii \cite{Sharma2024,Awais2025,Mustafa2024,Rani2024,Paul2025}. However, Singh \textit{et al}~\cite{singhclass1} demonstrated that for an isotropic embedding class-I spacetime, the only possible solutions in general relativity are either flat spacetimes or the Schwarzschild interior solution~\cite{schwarzschild1916} and the Kohler-Chao solution~\cite{kohler1965}. As linear $f(Q)$ gravity is dynamically equivalent to general relativity (GR), the same restriction applies in the case of a linear $f(Q)$ gravity model solved via Karmarkar's condition. One is thus tempted to assume an anisotropic system to generate more realistic class-I solutions. This requirement directs attention to specific metric choices that can serve as viable seeds. In our study, we consider the Vaidya-Tikekar metric ansatz, which has found many applications in the modelling of compact relativistic stars \cite{vaidya}. It is noteworthy that in the Vaidya-Tikekar (VT) geometry, the $t=\text{constant}$ hypersurface can be embedded in a four-dimensional Euclidean space as a spheroidal geometry rather than a spherical geometry \cite{vaidya}. This observation leads to a specific form of the radial metric potential that allows exact interior solutions to be obtained without assuming an equation of state (EOS). Solutions constructed by this method produce physically acceptable profiles for ultra-dense stellar interiors. Subsequently, the VT ansatz has become a widely used tool in relativistic stellar structure. The Vaidya-Tikekar (VT) ansatz has also been extensively used to construct anisotropic stellar models~\cite{maurya19vt,maurya22vt,karmarkar0,paul,Sharma2012}, to derive mass and compactness bounds~\cite{Sharma2006}, higher dimensional studies~\cite{Khugaev2016,Chanda2024} and charged solutions corresponding to the exterior consistent with Reissner-Nordström geometry~\cite{koma,kumar2014,chatto,Sharma2021}. Very recently, the ansatz has been used to model strange quark stars and anisotropic embedding class-I solutions within $f(Q)$ gravity~\cite{Kaur2024b,Sharma2024,Awais2025}. These developments show that the VT geometry remains a flexible and reliable basis for modelling dense stellar interiors in both general relativity and its extended versions. Earlier, VT embedding class-I anisotropic models generated via gravitational decoupling were constructed in GR~\cite{maurya2022ps}, where physically viable compact objects were successfully developed. 
The present work combines gravitational decoupling with embedding class-I geometry in linear $f(Q)$ gravity to construct a generalized class of compact star models. In GR-based decoupling models, the only deformation freedom arises from the decoupling parameter, which modifies the metric components. Any modification in mass or compactness is inseparably tied to this geometric deformation. There is no independent mechanism within GR to distinguish between effects arising from geometric deformation and those associated with matter-sector normalization. GR-based decoupling cannot distinguish whether high mass compact objects arise from geometric stiffening or from modified gravitational coupling. In contrast, a linear $f(Q)$ modification of the form $f(Q)=\beta_1 Q+\beta_2$ is dynamically equivalent to GR at the geometric level. Nevertheless, it introduces a gravitational coupling parameter $\beta_1$ that uniformly rescales the effective matter sector without altering the metric structure. When gravitational decoupling is implemented within this framework, the resulting system contains two independent parameters: the decoupling parameter $\epsilon$, which governs geometric deformation and EOS stiffness, and the coupling parameter $\beta_1$, which governs matter-sector rescaling as demonstrated below.
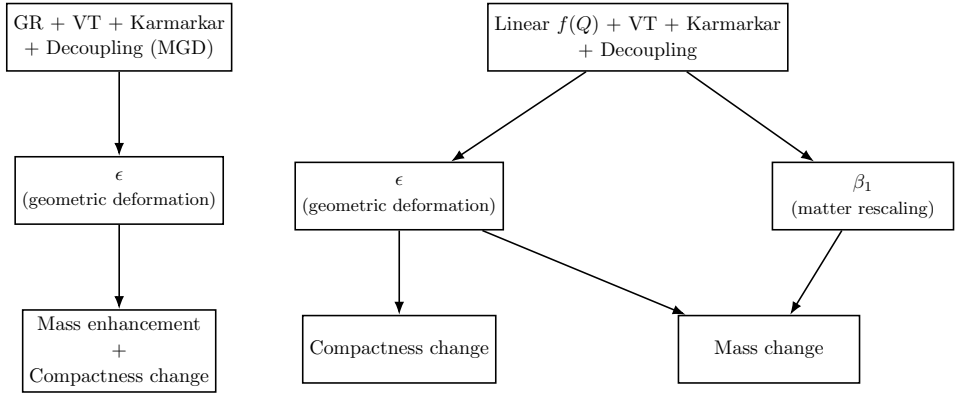
\begin{figure}[h]
\centering
\resizebox{\linewidth}{!}{
\begin{tikzpicture}[
    box/.style={draw, thick, align=center, minimum width=3.2cm, minimum height=1.2cm},
    arrow/.style={-Latex, thick},
    node distance=1.5cm and 2.8cm
]

\node[box] (gr) {GR + VT + Karmarkar \\ + Decoupling (MGD)};
\node[box, below=of gr] (eps1) {$\epsilon$ \\ \small (geometric deformation)};
\node[box, below=of eps1] (mass1) {Mass enhancement \\ $+$\\Compactness change};

\draw[arrow] (gr) -- (eps1);
\draw[arrow] (eps1) -- (mass1);

\node[box, right=4.5cm of gr] (fq) {Linear $f(Q)$ + VT + Karmarkar \\ + Decoupling};

\node[box, below left=1.6cm and -0.3cm of fq] (eps2) {$\epsilon$ \\ \small (geometric deformation)};
\node[box, below right=1.6cm and -0.3cm of fq] (beta) {$\beta_1$ \\ \small (matter rescaling)};

\node[box, below=of eps2] (comp) {Compactness change};
\node[box, right=3.2cm of comp] (mass2) {Mass change};

\draw[arrow] (fq) -- (eps2);
\draw[arrow] (fq) -- (beta);

\draw[arrow] (eps2) -- (comp);
\draw[arrow] (eps2) -- (mass2);

\draw[arrow] (beta) -- (mass2);

\end{tikzpicture}
}
\caption{
Structural distinction between GR-based decoupling and linear $f(Q)$ decoupling.
In GR, mass and compactness changes arise solely from geometric deformation $\epsilon$.
In linear $f(Q)$ gravity, $\epsilon$ modifies geometry (and hence compactness),
while $\beta_1$ independently shifts the mass scale without affecting compactness.
}
\end{figure}

The two-parameter framework $(\epsilon,~\beta_1)$ constitutes the central novelty of the present work, enabling a controlled comparison between GR and linear $f(Q)$ gravity at fixed geometric deformation (i.e., fixed $\epsilon$), and hence for fixed EOS stiffness, thereby isolating the role of matter-sector rescaling under identical geometric conditions. Such a separation cannot be achieved in GR alone, where geometric deformation and mass scaling are intrinsically linked.

The simultaneous presence of the geometric deformation parameter $\epsilon$ and the coupling parameter $\beta_1$ leads to a cumulative enhancement of the stellar mass that cannot be attained within either pure linear $f(Q)$ gravity or GR-based decoupling taken separately. While $\epsilon$ modifies the spacetime geometry and effectively controls the stiffness of the configuration, $\beta_1$ independently rescales the matter sector, thereby shifting the overall mass scale. Their combined action therefore produces a larger mass amplification than that obtained in single-parameter frameworks, where only geometric deformation is available.

The present work, thus, enlarges the deformation space of embedding class-I VT models from a single-parameter geometric framework (GR + decoupling) to a genuinely two-parameter framework $(\epsilon,~\beta_1)$. It is important to note that $\beta_1$ is a fundamental gravitational coupling constant of the theory, and it should remain universal, i.e., the same for all stellar configurations, as it characterizes the underlying gravity theory rather than individual objects. Consequently, variations in stellar properties at fixed $\beta_1$ reflect genuine geometric effects, whereas comparisons at fixed $\epsilon$ isolate gravitational coupling-induced mass shifts within the same geometric background. This interplay between $\epsilon$ and $\beta_1$ enables access to higher-mass configurations, including those approaching or populating the neutron star-black hole mass-gap region, within a physically consistent and controlled framework that is not achievable in GR-based decoupling models.

The structure of the paper is as follows. In Sec.~\ref{sec2}, we derive the field equations for linear $f(Q)$ gravity in an embedding class-I spacetime in the presence of an additional source. In Sec.~\ref{sec3}, we implement minimal geometric deformation using the Vaidya-Tikekar metric ansatz to obtain a modified class-I anisotropic solution. In Sect.~\ref{sec4}, we match the interior configuration to the exterior Schwarzschild/Schwarzschild-(anti-)de Sitter spacetime and determine the model parameters. In Sec.~\ref{sec5}, we analyze physically interesting quantities such as energy density, pressures, anisotropy, compactness, EOS and mass-radius relations of the resultant stellar configuration. We also compare the estimated radii of some pulsars in this construction with observational data. In Sec.~\ref{sec6}, we conclude by summarizing the key findings of our investigation.

\section{The field equations}\label{sec2}

The action in $f(Q)$ gravity can be extended by introducing an additional source term described by the Lagrangian $L_{\theta}$, coupled through a constant $\epsilon$. With both the matter and auxiliary sectors included, the full action takes the form \cite{maurya2023cf}
\begin{equation}
S=\left( \int \frac{1}{2} f(Q)\sqrt{-g}\, d^4x + \int L_m \sqrt{-g}\, d^4x \right)
+ \epsilon \int L_{\theta} \sqrt{-g}\, d^4x .
\end{equation}

Here $L_m$ represents the matter Lagrangian, producing the energy-momentum tensor $T_{\mu\nu}$ in $f(Q)$ gravity, where the dynamics follow from the non-metricity scalar $Q$. Units are chosen so that $\frac{8\pi G}{c^4}=1$. The additional source term $L_{\theta}$ contributes to the physical behaviour of the system beyond pure $f(Q)$ models.

The stress tensors associated with the two sectors are
\begin{equation}
T{\mu}{}{\nu} = -\frac{2}{\sqrt{-g}}
\frac{\delta(\sqrt{-g}L_m)}{\delta g^{\mu\nu}},
\end{equation}
\begin{equation}
T_{\mu\nu}{}^{\theta} = -\frac{2}{\sqrt{-g}}
\frac{\delta(\sqrt{-g}L_{\theta})}{\delta g^{\mu\nu}}.
\end{equation}
Combining these two contributions, the effective matter source becomes
\begin{equation}
\bar{T}_{\mu\nu}=T_{\mu\nu}+ \epsilon\, T_{\mu\nu}{}^{\theta},
\end{equation}
and conservation of the total energy-momentum tensor follows from the Bianchi identity:
\begin{equation}
\nabla_{\mu}\bar{T}_{\mu\nu}=0.
\end{equation}

The non-metricity tensor is defined through the affine connections
\begin{equation}
Q_{\lambda\mu\nu}=\nabla_{\lambda} g_{\mu\nu}
=\partial_{\lambda}g_{\mu\nu}
-\Gamma^{\delta}{}_{\lambda\mu} g_{\delta\nu}
-\Gamma^{\delta}{}_{\lambda\nu} g_{\mu\delta}.
\end{equation} 

The general connection is written as
\begin{equation}
\Gamma^{\delta}{}_{\mu\nu}
= \{^{\delta}{}_{\mu\nu}\}
+ K^{\delta}{}_{\mu\nu}
+ L^{\delta}{}_{\mu\nu},
\end{equation}
where the respective Levi-Civita connection, contortion and disformation are respectively given by 
\begin{align}
\{^{\delta}{}_{\mu\nu}\}
&=\frac12 g^{\delta\sigma}
(\partial_{\mu}g_{\sigma\nu}+\partial_{\nu}g_{\sigma\mu}-\partial_{\sigma}g_{\mu\nu}),\\[4pt]
L^{\delta}{}_{\mu\nu}
&=\frac12Q^{\delta}{}_{\mu\nu}
 - Q_{(\mu}{}^{\delta}{}_{\nu)},\\[4pt]
K^{\rho}{}_{\mu\nu}
&=\frac12T^{\delta}{}_{\mu\nu}
+ T_{(\mu}{}^{\delta}{}_{\nu)},
\end{align}
with $T^{\delta}{}_{\mu\nu}=2\Gamma^{\delta}{}_{[\mu\nu]}$ as the torsion tensor. Subsequently, the super-potential is introduced as
\begin{equation}
P^{\alpha}{}_{\mu\nu}
=\frac14\left(
 -Q^{\alpha}{}_{\mu\nu}
 +2 Q^{\alpha}{}_{(\mu\nu)}
 +Q^{\alpha} g_{\mu\nu}
 -\tilde{Q}^{\alpha} g_{\mu\nu}
 -\delta^{\alpha}{}_{(\mu} Q_{\nu)}
\right),
\end{equation}
where the traces
\begin{equation}
Q_{\alpha}=Q_{\alpha}{}^{\mu}{}_{\mu},
\qquad
\tilde{Q}_{\alpha}=Q^{\mu}{}_{\mu\alpha},
\end{equation}
allow the non-metricity scalar to be written as
\begin{equation}
Q=-Q_{\alpha\mu\nu}P^{\alpha\mu\nu}.
\end{equation}
Variation of the action with respect to the metric yields the field equations
\begin{equation}
\frac{2}{\sqrt{-g}}\,
\nabla_{\gamma}\!\left(\sqrt{-g}\, f_Q P^{\gamma}{}_{\mu\nu}\right)
+\frac12 g_{\mu\nu} f
+ f_Q \Big(
P_{\mu\gamma\delta} Q_{\nu}{}^{\gamma\delta}
- 2 Q_{\gamma\delta\mu} P_{\nu}{}^{\gamma\delta}
\Big)
= -\bar{T}_{\mu\nu},
\end{equation}
and variation with respect to the connection leads to the condition
\begin{equation}
\nabla_{\mu}\nabla_{\nu}
\big(\sqrt{-g}\,f_Q P^{\gamma}{}_{\mu\nu}\big)=0,
\end{equation}
where $f_Q=df/dQ$.

In the absence of torsion and curvature, the affine parameters get fixed to a pure gauge form
\begin{equation}
\Gamma^{\lambda}{}_{\mu\nu}
=\frac{\partial x^{\lambda}}{\partial \xi^{\beta}}
\frac{\partial^2 \xi^{\beta}}{\partial x^{\mu}\partial x^{\nu}},
\end{equation}
which, in the coincident gauge, vanishes i.e.,
\begin{equation}
\Gamma^{\lambda}{}_{\mu\nu}=0,
\end{equation}
Subsequently, the non-metricity tensor reduces to the simple form
\begin{equation}
Q_{\lambda\mu\nu}=\partial_{\lambda} g_{\mu\nu}.
\end{equation}

Having established the general structure of the theory, we now consider a static and spherically symmetric spacetime describing the interior of a relativistic star. In the present work, to obtain  an embedding class-I solution in $f(Q)$ gravity obtained through the gravitational decoupling method, we assume the line element in the form
\begin{equation}
     ds^{2} = -e^{\nu(r)}dt^{2}+e^{\lambda(r)}dr^{2}+r^{2}d\Omega^{2}, \label{im}
\end{equation}
where $\nu(r)$ and $\lambda(r)$ are two unknown metric potentials. The line-element considered here belongs, in general, to an embedding class II geometry. Karmarkar \cite{karmarkar} demonstrated that such a spacetime can be reduced to embedding class I by imposing a specific geometric constraint. This condition guarantees that a four-dimensional spacetime admits an isometric embedding in a five-dimensional flat Euclidean manifold, provided the relation
\begin{equation}
R_{1414} =
\frac{R_{1212} R_{3434} + R_{1224} R_{1334}}
{R_{2323}}, \label{kc1}
\end{equation} holds with $R_{2323} \ne 0$. Using the line element in Eq.~(\ref{im}), the non-vanishing Riemann tensor components take the form
\begin{eqnarray}
R_{1414} &=& -e^{\nu}\left( \frac{\nu''}{2} + \frac{\nu'^2}{4} - \frac{\lambda'\nu'}{4} \right),\label{r1}\\[4pt]
R_{2323} &=& -e^{\lambda}\,r^{2}\sin^{2}\theta\,\big(e^{\lambda} - 1\big),\label{r2}\\[4pt]
R_{1334} &=& R_{1224}\,\sin^{2}\theta = 0,\label{r3}\\[4pt]
R_{1212} &=& \frac{1}{2}\,r\,\lambda', \label{r4}\\[4pt]
R_{3434} &=&-\frac{1}{2}\,r\,\sin^{2}\theta\,\nu'\,e^{\nu-\lambda}. \label{r5}
\end{eqnarray}
Using Eq.~(\ref{kc1})-(\ref{r5}), we obtain one of the metric potentials 
\begin{equation}
 e^{\nu} = \left( C\int\sqrt{e^{\lambda}-1}\,dr + D \right)^2,  \label{metricsk} 
\end{equation} 
in terms of a single generating function $\lambda(r)$. In (\ref{metricsk}), $C$ and $D$ are integration constants which can be fixed by the appropriate junction conditions.

The non-metricity scalar corresponding to this spacetime then follows as
\begin{equation}
    Q = -\frac{2 e^{-\lambda}(\nu' r+1)}{ r^2}. \label{q}
\end{equation}

We consider the internal composition of the self-gravitating system in pure $f(Q)$ gravity described by an anisotropic fluid distribution. This choice is consistent with the geometric character of embedding class-I spacetime, where the metric potentials are constrained by the Karmarkar condition. As noted in the analysis of Singh \textit{et al}~\cite{singhclass1}, the Karmarkar condition in linear $f(Q)$ gravity severely limits the admissible isotropic stellar solutions. One can, however, generate a new class of solutions within class-I geometry that admit a linear form of the function $f(Q)$, provided an additional degree of freedom is incorporated into the system of equations. We introduce this additional degree of freedom by choosing the matter distribution to be anisotropic in nature, which is a reasonable choice in the context of compact stellar objects. Consequently, we write the energy-momentum tensor $T_{\mu{\nu}} $ in the form
\begin{equation}
T_{\mu \nu}
   = (\rho + p_t)\,u_\mu u_\nu
     + p_t g_{\mu \nu}
     + (p_r - p_t)\, v_\mu v_\nu ,\label{tmatter2}
\end{equation}
where $\rho$, $p_r$ and $p_t$ are the energy density, radial and transverse pressure, respectively. We also denote the components of the energy-momentum tensor of the new source as
\begin{eqnarray}
\left[T^\theta\right]^0_{\;0} &=& -\rho^\theta , \\
\left[T^\theta\right]^1_{\;1} &=& p^\theta_r , \\
\left[T^\theta\right]^2_{\;2} &=& \left[T^\theta\right]^3_{\;3} = p^\theta_t.
\end{eqnarray} 
Subsequently, the total energy density and the two pressures take the form
\begin{eqnarray}
\rho_{\rm}^{ tot} &=& \rho + \varepsilon \, \rho^\theta , \label{rhogd} \\
P_r^{\rm tot} &=& p_r + \varepsilon \, p_r^\theta , \label{prgd}\\ 
P_t^{\rm tot} &=& p_t + \varepsilon \, p_t^\theta .\label{ptgd}
\end{eqnarray}
The field equations for line element (\ref{im}) take the form
\begin{eqnarray}
\rho^{tot}&=&{f_Q} \left(\frac{{\lambda}'+\nu'}{e^\lambda r}+Q+\frac{1}{r^2}\right)-\frac{f}{2}, \label{rho}\\
P_r^{tot}&=&-{f_Q} \left(Q+\frac{1}{r^2}\right)+\frac{f}{2}, \label{pr}\\
P_t^{tot}&=& -{f_Q} \left(\frac{Q}{2}-e^\lambda\left[(\frac{\nu'}{4}+\frac{1}{2r})(\nu'-\lambda')+\frac{\nu''}{2}\right]\right)+\frac{f}{2}, \label{pt}\\
0&=&\frac{\cot\theta}{2} Q' f_{QQ}, \label{fqq}
\end{eqnarray}
where $\Delta =P_t^{tot}-P_r^{tot}=(p_t-p_r)+\epsilon(p_t^{\theta}-p_r^{\theta})$, is the measure of anisotropy.

Before specifying the explicit form of $f(Q)$, it is important to recall that 
the field equations of STGR impose non-trivial restrictions on the admissible Lagrangian. Wang \textit{et al}~\cite{Wang2022} showed that, in the coincident gauge, the off-diagonal component of the field equations forces either $f_{QQ}=0$ or $Q' = 0$. The second branch corresponds to a constant non-metricity scalar and leads to interior Schwarzschild-type solutions that need not be asymptotically flat and, hence, cannot be treated as stellar vacuum solutions. The viable branch is, therefore, the one with $f_{QQ}=0$, which uniquely fixes a linear form of $f(Q)$. The choice of a linear form of $f(Q)$ is not an assumption; it is the only functional form consistent with the physically acceptable vacuum solution. Consequently, we assume
\begin{equation}
    f(Q)=\beta_1Q+\beta_2, \label{fq}
\end{equation} where, $\beta_1$ and $\beta_2$ are $f(Q)$ gravity parameters.
Using Eq.~(\ref{im}) and (\ref{fq}), Eq.~(\ref{rho})-(\ref{pt}) take the form
\begin{eqnarray}
\rho^{tot}&=& \frac{1}{2 r^{2}}
\left[
2\beta_{1}
+ 2 e^{-\lambda}\,\beta_{1}\,(r\nu' - 1)
- r^{2}\beta_{2}
\right],
\label{rho1}
\\
P^{tot}_r&=& \frac{1}{2 r^{2}}
\left[
-2\beta_{1}
+ 2 e^{-\lambda}\,\beta_{1}\,(r\nu' + 1)
+ r^{2}\beta_{2}
\right],
\label{pr1}\\
P_{t}^{tot} &=& \frac{e^{-\lambda}}{4 r}
\left[
2 e^{\lambda} r\beta_{2}
+ \beta_{1}(2 + r\nu')(\nu'-\lambda')
+ 2 r \beta_{1} \nu''
\right].
\label{pt1}
\end{eqnarray}
With the field equations established, the following section applies the 
gravitational decoupling (GD) procedure to split the system into two 
independent sectors while preserving the embedding class-I structure.

\section{Field equations using MGD and Karmarkar's condition} \label{sec3}
In this section, we analyze the field equations of $f(Q)$ gravity within the framework of minimal geometric deformation (MGD). Gravitational decoupling is
introduced through linear deformations of the metric potentials. The radial
sector is modified as
\begin{equation}
    e^{-\lambda(r)} = W(r) + \epsilon\,\psi(r),\label{rad}
\end{equation}
where $W(r)$ denotes the seed metric function in pure $f(Q)$ gravity, and
$\psi(r)$ represents the radial geometric deformation.
Following the standard MGD approach, the temporal potential is also written
as a linear deformation
\begin{equation}
    \nu(r) = H(r) + \epsilon\,\eta(r), \label{temp}
\end{equation}
with $H(r)$ taken as the seed temporal metric and $\eta(r)$ as the induced temporal deformation. When the Karmarkar condition is imposed, the
function $\eta(r)$ is determined through the embedding class-I relation, since $\nu(r)$ becomes coupled to the deformed radial metric. Thus, while the temporal
deformation remains linear in $\epsilon$, its explicit form is fixed non-trivially by the class-I constraint.
The MGD formalism allows the full set of decoupled field equations in $f(Q)$ gravity to be reorganized into two mutually independent sectors. The first sector corresponds to the seed configuration sourced by $T_{\mu\nu}$ and is described entirely within pure $f(Q)$ gravity. The second sector contains the
contributions generated by the additional source $\theta_{\nu}$ and depends solely on the deformation functions.

\subsection{Seed system in pure $f(Q)$ gravity ($\epsilon=0$)}
For the undeformed sector, the field equations reduce to
\begin{equation}
    \rho = \frac{\beta_{1}(1-W)}{r^{2}}
           - \frac{\beta_{1} W'}{r}
           - \frac{\beta_{2}}{2},
    \label{rho_seed}
\end{equation}
\begin{equation}
    p_r = \frac{\beta_{1}(W-1)}{r^{2}}
        + \frac{\beta_{1} H' W}{r}
        + \frac{\beta_{2}}{2},
    \label{pr_seed}
\end{equation}
\begin{equation}
    p_t = 
    \frac{\beta_{1}(W'H' + 2H''W + H'^2 W)}{4}
    + \frac{\beta_{1}(W' + H'W)}{2r}
    + \frac{\beta_{2}}{2}.
    \label{pt_seed} 
\end{equation}
For the seed solution, we choose the embedding class-I Vaidya-Tikekar metric ansatz \cite{vaidya}
\begin{equation}
W^{-1}(r) =
\frac{1 + K\,r^{2}/L^{2}}{1 - r^{2}/L^{2}}, \label{elamv}
\end{equation}
where $K$ and $L$ are curvature parameters of the associated spacetime. In Eq.~(\ref{metricsk}), this assumption readily provides the unknown metric potential 
\begin{equation}
e^{H(r)} =
\left[\, C - D\,\sqrt{(1+K)\,\left(L^{2}-r^{2}\right)}\, \right]^{2}.\label{enuv}
\end{equation}
    
\subsection{System with the additional source $\theta_{\mu\nu}$ ($\epsilon\ne0$)}
In the presence of the additional source, we obtain the following equations for the deformation sector 
\begin{equation}
    \rho^{\theta}
      = -\beta_{1}\!\left(
        \frac{\psi}{r^{2}}
        + \frac{\psi'}{r}
        \right),
    \label{rho_theta}
\end{equation}
\begin{equation}
    p^{\theta}_{r}
      = \beta_{1}\!\left(
        \frac{\psi}{r^{2}}
        + \frac{\nu'\psi}{r}
        + \frac{W \eta'}{r}
        \right),
    \label{pr_theta}
\end{equation}
\begin{equation}
\begin{aligned}
    p^{\theta}_{t}
     &= \beta_{1}\!\bigg[
        \frac{\psi'\nu'}{4}
        + \frac{\nu''\psi}{2}
        + \frac{\nu'^2\psi}{4}
        + \frac{\psi'}{2r}
        + \frac{\nu'\psi}{2r}
        \bigg] \\
     &\quad
       + \beta_{1}\bigg[ \frac{W}{4}
        \left( 2\eta'' + \beta_1\eta'^2
               + \frac{2\eta'}{r}+2H'\eta'\right)+\frac{W'\eta'}{4}\bigg].             
\end{aligned}
\label{pt_theta}
\end{equation}
As the seed sector is fully specified, we now have a system containing five
unknown functions, namely $\psi(r)$, $\eta(r)$, $\rho^{\theta}$, $p_{r}^{\theta}$,
and $p_{t}^{\theta}$. Along with the Karmarkar condition, we now have three independent equations and hence, we need to impose an additional constraint to close the system. In this context, some of the well-known techniques adopted so far by various investigators are the following:  
(i) the \textit{mimicking density condition}, in which the density of the source sector follows that of the seed i.e, $\rho =\rho^\theta$~\cite{maurya2020epjcnon};  
(ii) the \textit{mimicking pressure condition}, where the radial pressure of the source reproduces the radial pressure of the seed i.e., $p_{r} =p_r^{\theta}$~\cite{maurya2020epjcnon}; and  
(iii) the \textit{mimicking mass condition}, where the mass profile of the seed is identified with that of the source i.e., $m = m^{\theta}$~\cite{maurya2024decouple}.  

In the present analysis, we employ the \textit{mimicking density condition}, as it provides the most direct way to close the system while keeping the effective density well behaved. Moreover, it allows the deformation to be determined from a local relation, without imposing additional constraints on the pressure or the mass function. This makes it the simplest and least restrictive option for the present model. By applying the mimicking density condition and making use of
Eq.~(\ref{rho_seed}) and (\ref{rho_theta}), we obtain the following first-order
differential equation for the deformation function $\psi(r)$ 
\begin{equation}
    \psi'(r)
    + \frac{\psi(r)}{r}
    - \frac{\beta_{2}\,r}{2\beta_{1}}
    + \frac{1 - W(r)}{r}
    - W'(r)
    = 0.
    \label{diff}
\end{equation}
Solution of Eq.~(\ref{diff}) is obtained in the form
\begin{equation}
    \psi(r)
    = \frac{r^{2}}{6\beta_{1}}
      \left(
        \beta_{2}
        - \frac{6\beta_{1}(K+1)}{K r^{2} + L^{2}}
      \right)
      + \frac{c_{1}}{r},
    \label{psir}
\end{equation}
where $c_{1}$ is an integration constant. Regularity of the function at the stellar centre
($r=0$) leads to $c_{1}=0$.

Substituting Eq.~(\ref{psir}) in Eq.~(\ref{rad}) and (\ref{elamv}), the
deformed radial metric component takes the form
\begin{equation}
    e^{\lambda}
    =
    \frac{
        6\beta_{1}\left(K r^{2} + L^{2}\right)
    }{
        \beta_{2}K r^{4}\,\epsilon
        - 6\beta_{1} r^{2}\left(K\epsilon + \epsilon + 1\right)
        + L^{2}\left(6\beta_{1} + \beta_{2} r^{2}\epsilon\right)
    }.
    \label{modelam}
\end{equation}
 We note that the parameter $\beta_{2}$ is associated with the cosmological constant $\Lambda$ and is expected to have a negligible effect on stellar modelling. Hence, it is reasonable to assume $\beta_2 \approx 0$. This choice is consistent with the exterior matching conditions (see Sect.~\ref{sec4}), where a vanishing or negligible $\beta_2$ ensures compatibility with a Schwarzschild-type exterior spacetime and reflects the negligible role of cosmological constant at stellar scales.  Under this assumption, the radial metric takes a simplified form
\begin{equation}
    e^{\lambda}
    =
    \frac{
        K r^{2} + L^{2}
    }{
        L^{2} - r^{2}\left(K\epsilon + \epsilon + 1\right)
    }.
    \label{modelam1}
\end{equation}

The corresponding temporal component follows from the Karmarkar embedding
condition Eq.~(\ref{metricsk}) and yields
\begin{equation}
    e^{\nu}
    =
    \left[
        C
        -
        \frac{
            D (K+1) (\epsilon + 1)
        }{
            (K\epsilon + \epsilon + 1)\,
            \sqrt{
                \dfrac{(K+1)(\epsilon+1)}{
                   L^{2}- r^{2}(K\epsilon + \epsilon + 1)
                }
            }
        }
    \right]^{2}.
    \label{modenu}
\end{equation}
It is interesting to note that the modified metric functions become independent of
the $f(Q)$ parameters $\beta_{1}$ and $\beta_{2}$ in this construction. This reinforces the fact that
$f(Q)$ gravity does not alter the underlying spacetime geometry relative to
general relativity. Nevertheless, when $\beta_{1}$ and $\beta_{2}
(\approx 0)$ appear in the expressions for density and pressure, the physical quantities are rescaled to their general relativistic values. 

Now, to completely specify the $\theta$ sector, we obtain the temporal deformation function using Eq.~(\ref{temp}) as
\begin{equation}
    \eta(r)=\frac{2}{\epsilon}\ln\left[\frac{ 
        C
        -
        \frac{
            D (K+1) (\epsilon + 1)
        }{
            (K\epsilon + \epsilon + 1)\,
            \sqrt{
                \dfrac{(K+1)(\epsilon+1)}{
                    L^2-r^{2}(K\epsilon + \epsilon + 1)
                }
            }
        }
    }{ C - D\,\sqrt{(1+K)\,\left(L^{2}-r^{2}\right)}}\right]. \label{eta}
\end{equation}
Consequently, we obtain the energy density and two pressures in the $\theta$ sector as
\begin{eqnarray}
\rho^\theta &=& 
\frac{\beta_1 (K+1)\left(K r^2 + 3L^2\right)}
{\left(K r^2 + L^2\right)^2},
\label{rhoth}
\\[12pt]
p_r^\theta &=&
\frac{\beta_1 (K+1)}{\epsilon (K r^2 + L^2)}
\Bigg[
\frac{
D (K+3)
- C \sqrt{\frac{K+1}{L^2 - r^2}}
}{
- C \sqrt{\frac{K+1}{L^2 - r^2}} + D(K+1)
}
\Bigg.
\nonumber\\[6pt] &&
\quad -
\Bigg.
\frac{
(\epsilon+1)\!\left[
D\!\left(3(K+1)\epsilon + K + 3\right)
- C (K\epsilon+\epsilon+1)
\sqrt{
-\frac{(K+1)(\epsilon+1)}
     {\,r^2(K\epsilon+\epsilon+1)-L^2\,}
}
\right]
}{
D (K+1)(\epsilon+1)
- C (K\epsilon+\epsilon+1)
\sqrt{
-\frac{(K+1)(\epsilon+1)}
     {\,r^2(K\epsilon+\epsilon+1)-L^2\,}
}
}
\Bigg],
\nonumber\\[-2pt] && \label{prth}
\\[12pt]
p_t^\theta &=&
\frac{\beta_1 (K+1)}{\epsilon (K r^2 + L^2)^2}
\Bigg[
\frac{
- C L^2 \sqrt{\frac{K+1}{L^2 - r^2}}
+ D (K+3)L^2
+ D K r^2
}{
- C \sqrt{\frac{K+1}{L^2 - r^2}}
+ D(K+1)
}
\Bigg.
\nonumber\\[6pt] &&
\quad
\Bigg.
-
\frac{
(\epsilon+1)\left[
- C L^2 (K\epsilon+\epsilon+1)
\sqrt{
-\frac{(K+1)(\epsilon+1)}
     {\,r^2(K\epsilon+\epsilon+1)-L^2\,}
}
\right.
}{%
D (K+1)(\epsilon+1)
- C (K\epsilon+\epsilon+1)
\sqrt{
-\frac{(K+1)(\epsilon+1)}
     {\,r^2(K\epsilon+\epsilon+1)-L^2\,}
}
}
\nonumber\\[6pt] &&
\qquad
\left.
+ D L^2\left(3(K+1)\epsilon + K + 3\right)
+ D K r^2 (K\epsilon+\epsilon+1)
\right]
\Bigg].
\label{ptth}
\end{eqnarray}

We thus have two distinct sectors - one describing the seed configuration and the other governing the deformation induced by $\theta_{\mu\nu}$. This demonstrates that the system is gravitationally decoupled. Subsequently, the effective energy density and pressures arise from the linear combination of both the sectors as given in Eq.~(\ref{rhogd})-(\ref{ptgd}), and we obtain the physical quantities as
\begin{eqnarray}
\rho^{tot} &=& 
\frac{\beta_1 (K+1) (\epsilon +1)\,\left(K r^2+3 L^2\right)}
{\left(K r^2+L^2\right)^2},
\label{rhof}
\\[10pt]
P_r^{tot} &=& 
\frac{\beta_1 (K+1) (\epsilon +1)}{K r^2+L^2}\,
\frac{\mathcal{N}_r}{\mathcal{D}},
\label{prf}
\\[10pt]
P_t^{tot} &=& 
\frac{\beta_1 (K+1) (\epsilon +1)}{\left(K r^2+L^2\right)^2}\,
\frac{\mathcal{N}_t}{\mathcal{D}},
\label{ptf}\\
\Delta&=& 
\frac{\beta_1 (K+1) (\epsilon +1)}{\left(K r^2+L^2\right)^2}\,
\frac{\mathcal{N}_\Delta}{\mathcal{D}},\label{deltaf}
\end{eqnarray}
where,
\begin{eqnarray}
\mathcal{D} &=& 
D (K+1) (\epsilon +1)
- C (K \epsilon +\epsilon +1)
\sqrt{\frac{(K+1) (\epsilon +1)}
{L^2-r^2 (K \epsilon +\epsilon +1)}},
\\[8pt]
\mathcal{N}_r &=& 
C (K \epsilon +\epsilon +1)
\sqrt{\frac{(K+1) (\epsilon +1)}
{L^2-r^2 (K \epsilon +\epsilon +1)}}
- D \left[3 (K+1)\epsilon +K+3\right],
\\[8pt]
\mathcal{N}_t &=& 
C L^2 (K \epsilon +\epsilon +1)
\sqrt{\frac{(K+1) (\epsilon +1)}
{L^2-r^2 (K \epsilon +\epsilon +1)}}
\nonumber\\
&&
-\, D L^2 (3 K\epsilon +K+3 \epsilon +3)
-\, D K r^2 (K \epsilon +\epsilon +1),\\
\mathcal{N}_\Delta&=& D (2 (K+1) \epsilon +K+2)-C (K \epsilon +\epsilon +1) \sqrt{\frac{(K+1) (\epsilon +1)}{L^2-r^2 (K \epsilon +\epsilon +1)}}.\label{D}
\end{eqnarray}

\section{Matching conditions}\label{sec4}
Having determined the complete interior geometry for the assumed linear form of $f(Q)$, we need to ensure that the resultant configuration represents a physically admissible compact star. This requires matching of the interior spacetime to the corresponding vacuum solution of $f(Q)$ gravity across the boundary $r=R$, where the matter pressure vanishes. The exterior vacuum geometry is described by the Schwarzschild-(anti-)de Sitter line element
\begin{equation}
    dS^2_{+}=-\bigg(1-\frac{2 M}{r}-\frac{\Lambda r^2}{3}\bigg) dt^2+\frac{dr^2}{\bigg(1-\frac{2 M}{r}-\frac{\Lambda r^2}{3}\bigg)}+ r^2(d\theta^2+\sin^2\theta d\phi^2), \label{extmetric}
\end{equation}
where $M$ and $\Lambda$ represent the total gravitational mass and the cosmological constant, respectively. Within the linear $f(Q)$ model characterized by the constants $\beta_{1}$ and $\beta_{2}$, the effective cosmological constant is obtained as $\Lambda = \beta_{2}/(2\beta_{1})$. Observational measurements indicate that the present value of $\Lambda$ is of the order of $10^{-122} \ \text{(Planck units)}$ or $\approx$ $10^{-52}\,\mathrm{m^{-2}}$ ~\cite{barrowshaw}, which is far too small to have any significant impact on the internal structure of compact stars. Consequently, the contribution of $\Lambda$ can be neglected in stellar configurations without affecting the physical behaviour of the model. Accordingly, we set $\beta_{2}=0$ in our studies. This effectively reduces the line element (\ref{extmetric} to the Schwarzschild exterior metric
\begin{equation}
    dS^2_{+}=-\bigg(1-\frac{2 M}{r}\bigg) dt^2+\frac{dr^2}{\bigg(1-\frac{2 M}{r}\bigg)}+ r^2(d\theta^2+\sin^2\theta d\phi^2). \label{schex}
\end{equation}
To determine the integration constants of the interior solution, we now impose the continuity of the metric potentials across $r=R$, together with the vanishing pressure condition at the boundary of the star: 
 by,\begin{eqnarray}
1 - \frac{2M}{R}  &=& e^{\nu(R)}, \label{match1} \\
1 - \frac{2M}{R}  &=& e^{-\lambda(R)}, \label{match2} \\
P_{r}^{tot}(R) &=& 0, \label{match3}
\end{eqnarray}
The above boundary conditions determine the constants as 
\begin{eqnarray}
    L&=&R\sqrt{\frac{-2 K M+(K+1)(1+\epsilon) R}{2M}},\label{L}\\
    C&=&\frac{\sqrt{M} (3 (K+1) \epsilon +K+3)}{2 \sqrt{R} (K \epsilon +\epsilon +1) \sqrt{\frac{M}{R-2 M}}},\label{c}\\
    D &=& \frac{\sqrt{M}}{\sqrt{2} R^{3/2}}, \label{d}
\end{eqnarray} 
in terms of total mass $M$, radius $R$ and the decoupling parameter $\epsilon$. Obviously, as the metric potentials are independent of the $f(Q)$ gravity parameters ($\beta_1,~\beta_2$), the constants are also independent of $\beta_1$ and $\beta_2$. In Eq.~(\ref{L}), for real values of $L$ we must have
\begin{equation}
    \epsilon > \frac{2KM}{R(1+K)}-1. \label{epsilon} 
\end{equation}
A more stringent bound on $\epsilon$ can be obtained from other physical requirements, such as the fulfilment of the causality condition. 

\section{Results and physical analysis}\label{sec5}
To assess the physical viability of our model, we consider the pulsar PSR~$J0614-3329$, whose gravitational mass and radius inferred from NICER data~\cite{Mauviard2025} are $M = 1.44\,M_\odot$ and $R = 10.29~\mathrm{km}$, respectively. Choosing these values as input parameters in our model, for a given curvature parameter $K=2$, the allowed range of the decoupling parameter $\epsilon$ is found to be $-0.725 < \epsilon < 1.7$, where we have used the constraint (\ref{epsilon}) and the causality condition. Within this admissible interval, the model constants corresponding to different values of $\epsilon$ (for $K=2$ and $\beta_1=0.9$) are tabulated in Table~\ref{tab1}.

\begin{table}[ht]
\tbl{Values of the model parameters for different choices of $\epsilon$ for the estimated mass and radius of the pulsar PSR~$J0614-3329$ ($M = 1.44~M_\odot$ and $R = 10.29~\mathrm{km}$) for a given curvature parameter (we assume $K=2$) with $\beta_1 = 0.9$ and $\beta_2 = 0$.\label{tab1}}
{\begin{tabular}{@{}cccc@{}} \toprule
$\epsilon$ & $L$ (km) & $C$ & $D$ \\
\colrule
$-0.2$ & 20.0947 & 3.06509 & 0.0312205 \\
$0$    & 23.6154 & 1.91568 & 0.0312205 \\
$0.1$  & 25.1919 & 1.73885 & 0.0312205 \\
$0.2$  & 26.6754 & 1.62833 & 0.0312205 \\
$0.5$  & 30.6988 & 1.45592 & 0.0312205 \\
$1$    & 36.4299 & 1.34098 & 0.0312205 \\
\botrule
\end{tabular}}
\end{table}

\begin{figure}[h]
\centering
        \begin{minipage}{0.48\textwidth}
			\centering
                {\bfseries \textbf{PSR J0614$-$3329}}
                \includegraphics[width=1\textwidth]{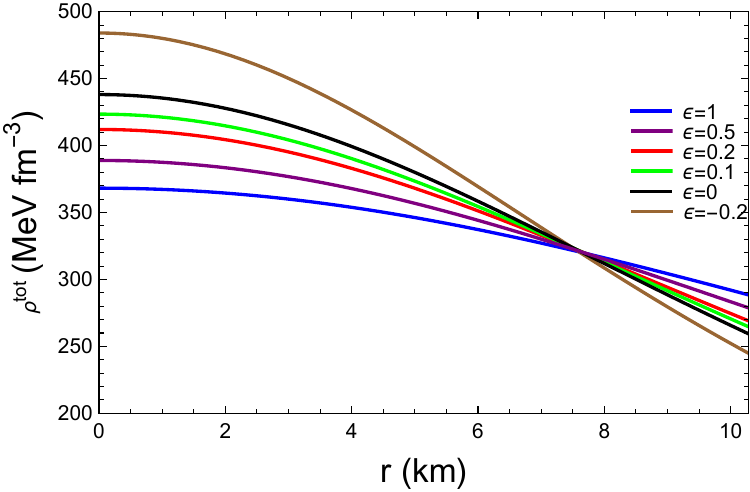}
                \caption{Radial variation of the total energy density $\rho^{tot}$ for for different values of $\epsilon$. We assume $K=2$,~$\beta_1 = 0.9$ and~$\beta_2=0$. }\label{fig1}
        \end{minipage}\hfill
        \vspace{0.5cm}
        \begin{minipage}{0.48\textwidth}
			\centering
                {\bfseries \textbf{PSR J0614$-$3329}}
                \includegraphics[width=1\textwidth]{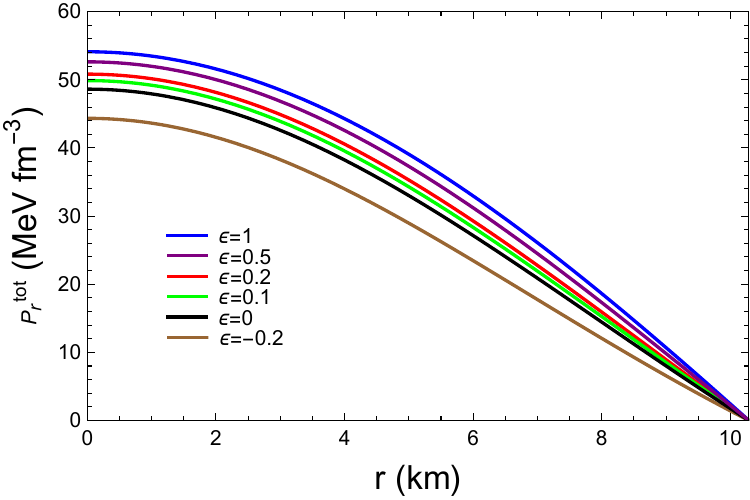}
                \caption{Radial variation of the total radial pressure $P_r^{tot}$ for different values of $\epsilon$. We assume $K=2$,~$\beta_1 = 0.9$ and~$\beta_2=0$.}\label{fig2}
        \end{minipage}\hfill
        \begin{minipage}{0.48\textwidth}
			\centering
                {\bfseries \textbf{PSR J0614$-$3329}}
                \includegraphics[width=1\textwidth]{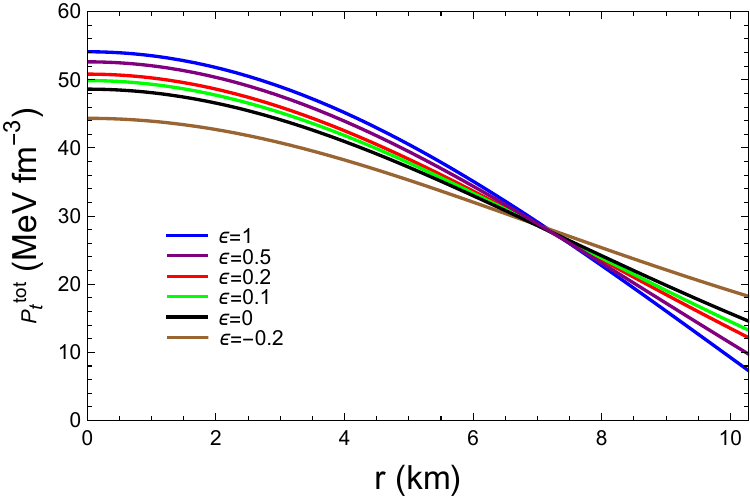}
                \caption{Radial variation of the total tangential pressure $P_t^{tot}$ for different values of $\epsilon$. We assume $K=2$,~$\beta_1 = 0.9$ and~$\beta_2=0$.}\label{fig3}
        \end{minipage}\hfill
          \vspace{0.5cm}
        \begin{minipage}{0.48\textwidth}
			\centering
                {\bfseries \textbf{PSR J0614$-$3329}}
                \includegraphics[width=1\textwidth]{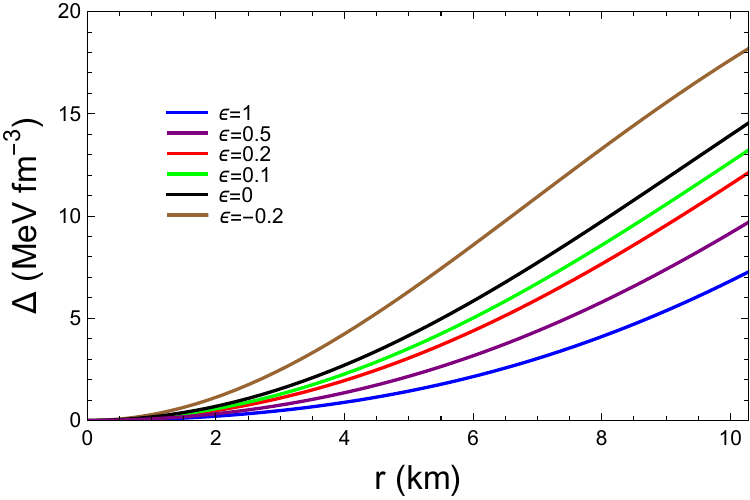}
                \caption{Radial variation of anisotropy $\Delta$ for different values of $\epsilon$. We assume $K=2$,~$\beta_1 = 0.9$ and~$\beta_2=0$.}\label{fig4}
        \end{minipage}\hfill
\end{figure}
\begin{figure}
         \begin{minipage}{0.48\textwidth}
			\centering
                {\bfseries \textbf{PSR J0614$-$3329}}
                \includegraphics[width=1\textwidth]{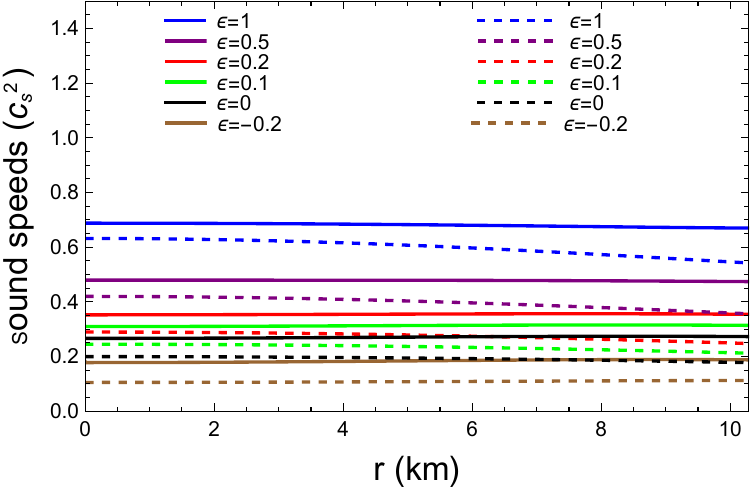}
                \caption{Radial variation of sound speeds ($c_s^2$) for different values of $\epsilon$. We assume $K=2$,~$\beta_1 = 0.9$ and~$\beta_2=0$. The radial and transverse sound speeds are denoted by solid and dashed lines, respectively.}\label{fig5}
        \end{minipage}\hfill
         \begin{minipage}{0.5\textwidth}
			\centering
                \includegraphics[width=1.1\textwidth]{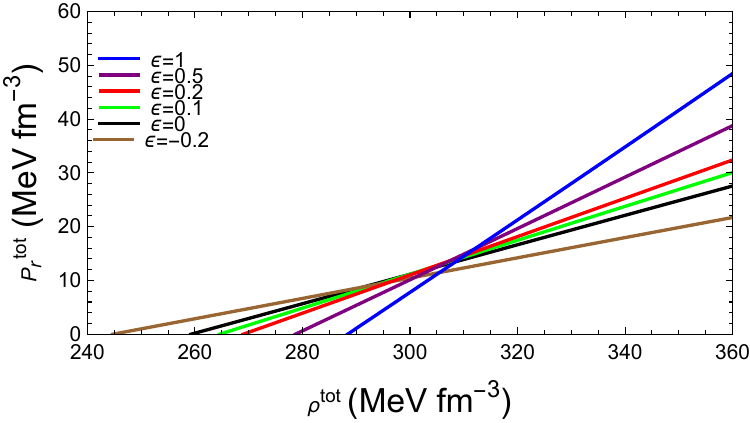}
                \caption{EOS of the matter distribution for different values of $\epsilon$. We assume $K=2$,~$\beta_1 = 0.9$ and~$\beta_2=0$.}\label{fig6}
        \end{minipage}\hfill         
\end{figure}

Profiles of the physically relevant quantities at the stellar interior, for a wide range of values, are shown in Fig.~\ref{fig1}-\ref{fig5}. The quantities are shown to be regular and well-behaved and capable of describing realistic compact stars such as PSR J0614$-$3329. Fulfilment of the physical acceptability conditions within the interior is analysed below:
\begin{itemize}

\item[(i)] The energy density, radial pressure and tangential pressure decrease 
monotonically from their central values toward the surface, as illustrated in 
Fig.~\ref{fig1}-\ref{fig3}.

\item[(ii)] The radial pressure is maximum at the centre and vanishes at 
the stellar boundary, as can be seen in Fig~\ref{fig2}.

\item[(iii)] The null energy condition (NEC), weak energy condition (WEC) and strong energy condition (SEC) are satisfied throughout the stellar interior as both the energy density and the two pressures remain positive throughout the stellar interior.

\item[(iv)] Fig.~\ref{fig1}-\ref{fig3} confirm that the dominant energy condition (DEC) ($\rho^{tot}_{c} \geq P^{tot}_{c}$) holds for all values of the parameter $\epsilon$.

\item[(v)] The causality condition $c_{s}^{2} = dP^{tot}/d\rho^{tot} \leq 1$ is satisfied at all radial points for the considered range of values of $\epsilon$, as shown in Fig.~\ref{fig5}.

\end{itemize}

In Fig.~\ref{fig1}-\ref{fig3}, we note that for a star of fixed mass and radius, up to a certain radial distance, the energy density decreases with increasing values of $\epsilon$, while both the radial and the tangential pressure increase with increasing values of $\epsilon$. Close to the crust region, this trend flips for energy density and transverse pressure while radial pressure increases with $\epsilon$ throughout the star. In Fig.~\ref{fig4}, we note that the anisotropy reduces with increasing values of the decoupling constant. 

It is interesting to note that, with mimicking energy density constraint $\rho=\rho^\theta$, the total density of the system becomes $\rho^{tot}=(1+\epsilon)\rho$ and hence, the total energy density might appear to be an increasing function of $\epsilon$. However, it should be stressed that, in the presence of a decoupling parameter, the model parameters $L$ and $C$ are also modified. In other words, different values of the decoupling parameter $\epsilon$ modify the geometry through  $L(\epsilon)$ and $C(\epsilon)$. Accordingly, the total energy density gets modified through $L(\epsilon)$ and the two pressures get modified through $L(\epsilon)$ as well as $C(\epsilon)$. As $\epsilon$ increases, the central energy density decreases to accommodate the fixed total mass ($M$)
\begin{equation}
    M=\frac{1}{2}\int_0^R \rho^{tot} r^2 dr. \label{mass}
\end{equation} 
In this study, with the total mass and radius fixed and a reduced density gradient in the inner region, the inner matter distribution provides comparatively less support against gravity. Consequently, the equilibrium condition in the presence of anisotropy demands a stronger radial pressure gradient. This justifies the monotonic increase of the total radial pressure with increasing values of $\epsilon$. The total tangential pressure is also sensitive to changes of $\epsilon$. In the core region, changes in $L$ and $C$ reinforce each other, leading to an increase in tangential pressure similar to that in the radial case. In the outer region, however, the density profile flattens, and the transverse stress gets modified accordingly to maintain equilibrium. This accounts for the attenuated or reversed growth of $P_t^{\rm tot}$ near the surface as $\epsilon$ increases. As the radial pressure increases throughout the stellar interior with increasing $\epsilon$ and the tangential pressure rises mainly in the core region and flips trend in the outer region, the difference $P_{t}^{\rm tot} - P_{r}^{\rm tot}$ decreases as $\epsilon$ increases. In effect, the overall anisotropy decreases for larger values of $\epsilon$. This reduction arises because the total mass redistribution induced by the modified geometry lowers the transverse stress required for equilibrium. This implies that stronger decoupling naturally drives the configuration toward greater isotropy.

The equation of state (EOS) $P_r^{\text{tot}} = P_r^{\text{tot}}(\rho^{\text{tot}})$, corresponding to different values of the decoupling parameter, is displayed in Fig.~\ref{fig6}. We note that the resultant EOS for all values of $\epsilon$ are nearly linear and hence, we perform a linear fit of the EOS in the form
\begin{equation}
	P_r^{\text{tot}} = \left(a\,\rho^{\text{tot}} - b\right),\label{eos}
\end{equation}
where the fitted coefficients $a$ and $b$ are listed in Table~\ref{tab2} for different values of $\epsilon$. It is interesting to note that for fixed values of $\beta_{1}$ and $K$, a stronger decoupling leads to a stiffer EOS.

\begin{table}[ht]
\tbl{EOS model parameters for different choices of $\epsilon$ for the estimated mass $M = 1.44~M_\odot$ and radius $R = 10.29~\mathrm{km}$ of the pulsar PSR J0614-3329. We assume $K=2$, $\beta_1 = 0.9$ and $\beta_2 = 0$.\label{tab2}}
{\begin{tabular}{@{}ccc@{}} \toprule
$\epsilon$ & $a$ & $b~(\mathrm{MeV\,fm^{-3}})$ \\
\colrule
$-0.2$ & 0.18507  & 45.061  \\
$0$    & 0.271845 & 70.3735 \\
$0.1$  & 0.313921 & 82.9916 \\
$0.2$  & 0.355468 & 95.5912 \\
$0.5$  & 0.47813  & 133.313 \\
$1$    & 0.679239 & 196.039 \\
\botrule
\end{tabular}}
\end{table}

We apply the fitted EOS to integrate the modified Tolman-Oppenheimer-Volkoff(TOV) equations
\begin{eqnarray}
\frac{d P_r^{\text{tot}}}{dr}
&=&
-\frac{\left(\rho^{\text{tot}} + P_r^{\text{tot}}\right)
\left[\, m^{eff}(r) + \frac{1}{2} r^3 P_r^{\text{tot}} \,\right]}
{r\left(r-2m^{eff}(r)\right)}
+ \frac{2}{r}\left(P_t^{\text{tot}} - P_r^{\text{tot}}\right),
\\[10pt]
\frac{dm^{eff}}{dr} &=& \frac{1}{2}\rho^{\text{tot}} r^2,
\end{eqnarray} 
to obtain the mass-radius ($M-R$) relationship for different choices of $\epsilon$.
\begin{figure}[h]
	\centering
	\includegraphics[width=1\textwidth]{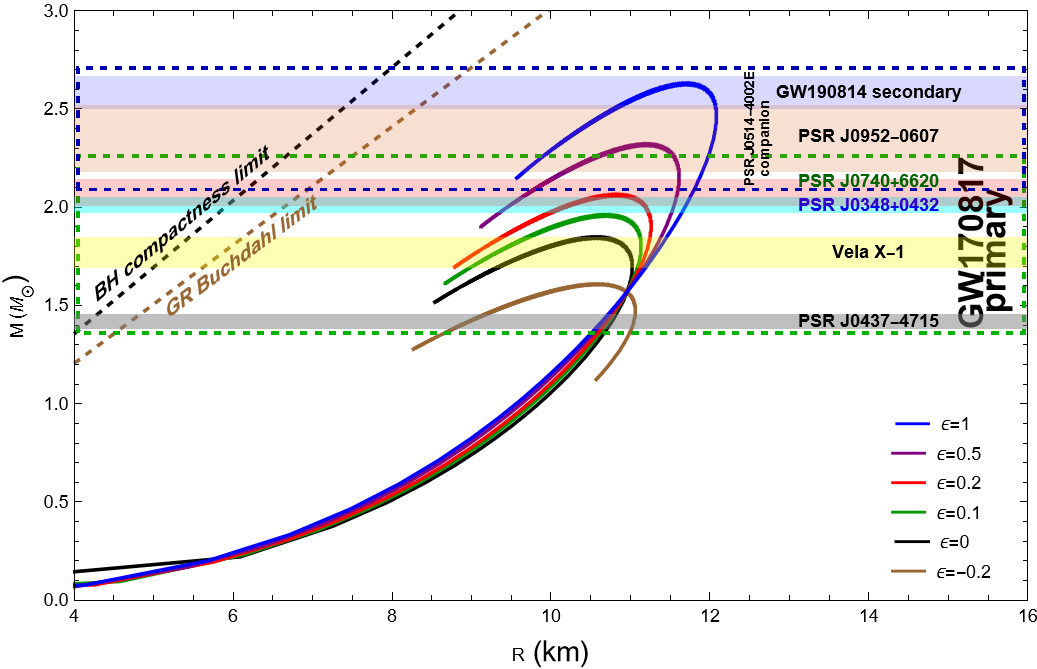}
	\caption{$M–R$ plots for different values of $\epsilon$ with $K=2$, $\beta_1=0.9$ and $\beta_2=0$.}
	\label{fig7}
\end{figure}
The results are shown in Fig.~\ref{fig7}. We note that a comparatively stronger decoupling yields a larger mass star. It is worth noting that for the assumed value of $\epsilon = -0.2$, the EOS becomes significantly softer, resulting in a very low surface density compared to other cases. Consequently, in the $M-R$ sequences, an incomplete curve is observed for the particular case of $\epsilon = -0.2$ as can be seen in Fig.~\ref{fig7}. Overall, we note that a comparatively stronger decoupling yields larger mass stars. This is an interesting observation as it demonstrates that gravitational decoupling introduces an additional degree of freedom through the parameter $\epsilon$, enabling finer control over the EOS and the resulting mass-radius profiles than is possible in linear $f(Q)$ gravity alone, which contains only the single gravitational parameter $\beta_{1}$.

\subsection{Comparison with GR}
\begin{figure}[h]
	\centering
	\begin{minipage}{0.48\textwidth}
		\centering
		{\bfseries \textbf{PSR J0614$-$3329}}
		\includegraphics[width=0.9\textwidth]{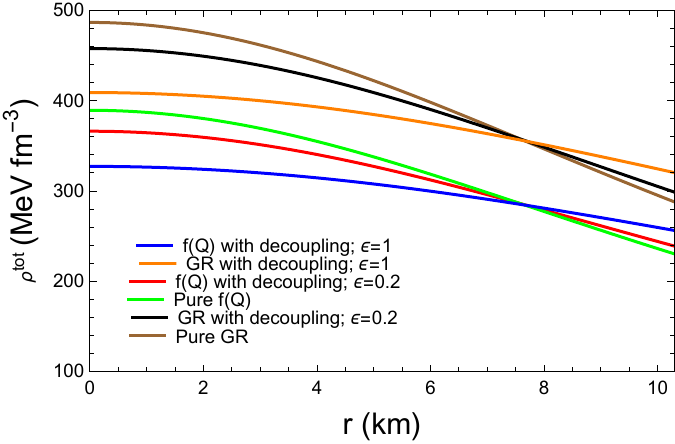}
		\caption{Radial variation of the total energy density $\rho^{tot}$ in GR and $f(Q)$ gravity (with $\beta_1$=0.8) in presence or absence of the decoupling parameter $\epsilon$ for assumed values of $K=2$ and~$\beta_2=0$.}\label{fig8}
	\end{minipage}\hfill
	\vspace{0.5cm}
	\begin{minipage}{0.48\textwidth}
		\centering
		{\bfseries \textbf{PSR J0614$-$3329}}
		\includegraphics[width=0.9\textwidth]{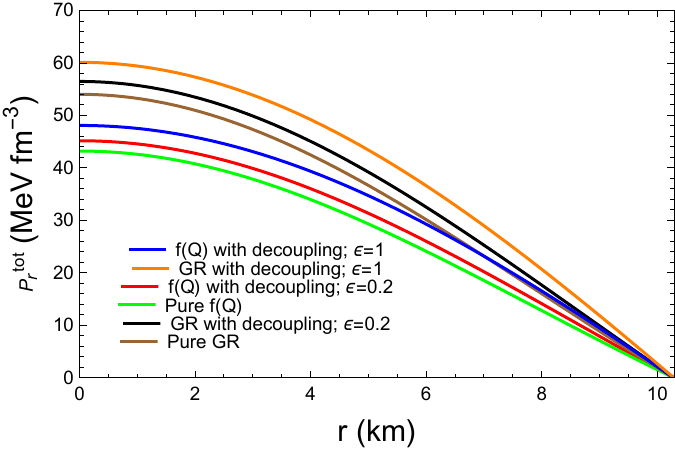}
		\caption{Radial variation of the total radial pressure $P_r^{tot}$ in GR and $f(Q)$ gravity (with $\beta_1$=0.8) in presence or absence of the decoupling parameter $\epsilon$ for assumed values of $K=2$ and~$\beta_2=0$.}\label{fig9}
	\end{minipage}\hfill
	\begin{minipage}{0.48\textwidth}
		\centering
		{\bfseries \textbf{PSR J0614$-$3329}}
		\includegraphics[width=0.9\textwidth]{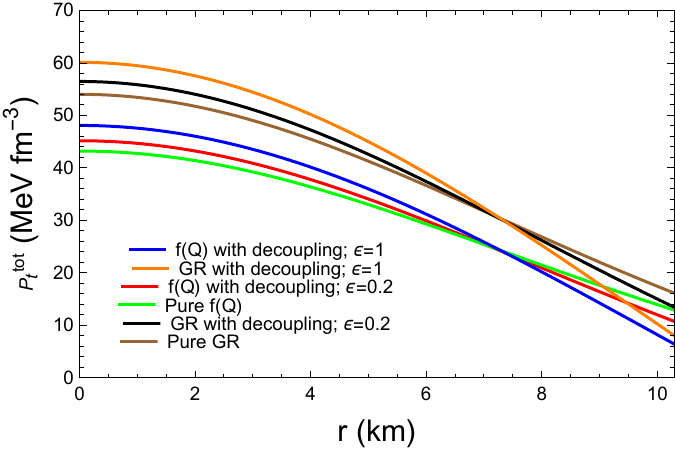}
		\caption{Radial variation of the total tangential pressure $P_t^{tot}$in GR and $f(Q)$ gravity (with $\beta_1$=0.8) in presence or absence of the decoupling parameter $\epsilon$ for assumed values of $K=2$ and~$\beta_2=0$.}\label{fig10}
	\end{minipage}\hfill
	\vspace{0.5cm}
	\begin{minipage}{0.48\textwidth}
		\centering
		{\bfseries \textbf{PSR J0614$-$3329}}
		\includegraphics[width=0.9\textwidth]{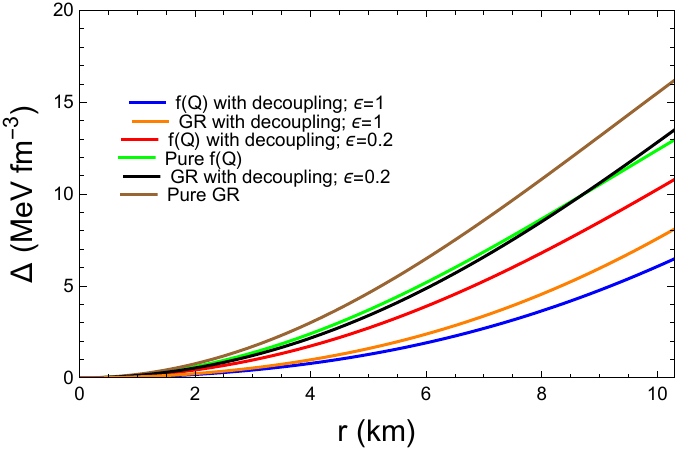}
		\caption{Radial variation of the anisotropy in GR and $f(Q)$ gravity (with $\beta_1$=0.8) in presence or absence of the decoupling parameter $\epsilon$ for assumed values of $K=2$ and~$\beta_2=0$.}\label{fig11}
	\end{minipage}\hfill
\end{figure}

\begin{figure}
	\begin{minipage}{0.48\textwidth}
		\centering
		{\bfseries \textbf{PSR J0614$-$3329}}
		\includegraphics[width=1\textwidth]{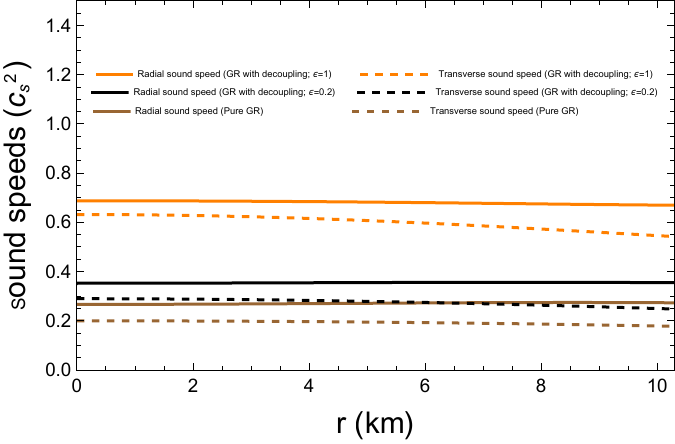}
		\caption{Radial variation of the sound speeds ($c_s^2$) in GR  in the presence or absence of the decoupling parameter $\epsilon$ for assumed values of $K=2$ and~$\beta_2=0$. The radial and transverse sound speeds are denoted by solid and dashed lines, respectively.}\label{fig12}
	\end{minipage}\hfill
	\vspace{0.5cm}
	\begin{minipage}{0.48\textwidth}
		\centering
		{\bfseries \textbf{PSR J0614$-$3329}}
		\includegraphics[width=1\textwidth]{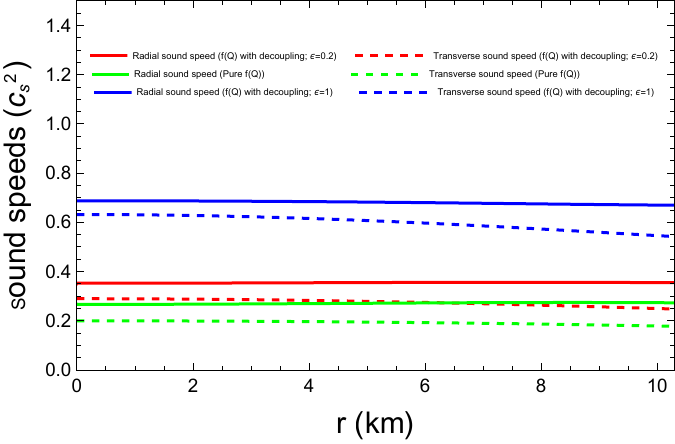}
		\caption{Radial variation of the sound speeds ($c_s^2$) in $f(Q)$ gravity (with $\beta_1$=0.8) in presence or absence of the decoupling parameter $\epsilon$ for assumed values of $K=2$ and~$\beta_2=0$. The radial and transverse sound speeds are denoted by solid and dashed lines, respectively.}\label{fig13}
	\end{minipage}\hfill
	\begin{minipage}{0.55\textwidth}
		\centering
		\includegraphics[width=1.1\textwidth]{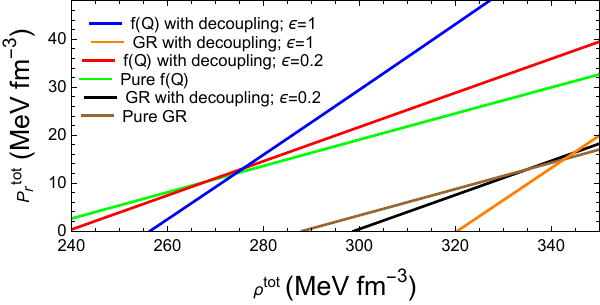}
		\caption{Variation of the EOS in GR and $f(Q)$ gravity (with $\beta_1$=0.8) in presence or absence of the decoupling parameter $\epsilon$ for assumed values of $K=2$ and~$\beta_2=0$.}\label{fig14}
	\end{minipage}\hfill
	\vspace{0.5cm}
\end{figure}

We now analyse our developed model by comparing the behaviour of the physical quantities in the presence and absence of the additional source term, both in GR and in $f(Q)$ gravity. The results are shown in Fig.~\ref{fig8}-\ref{fig14} and in Table~\ref{tab3} and \ref{tab4}.


\begin{figure}
	\centering
	\includegraphics[width=1\textwidth]{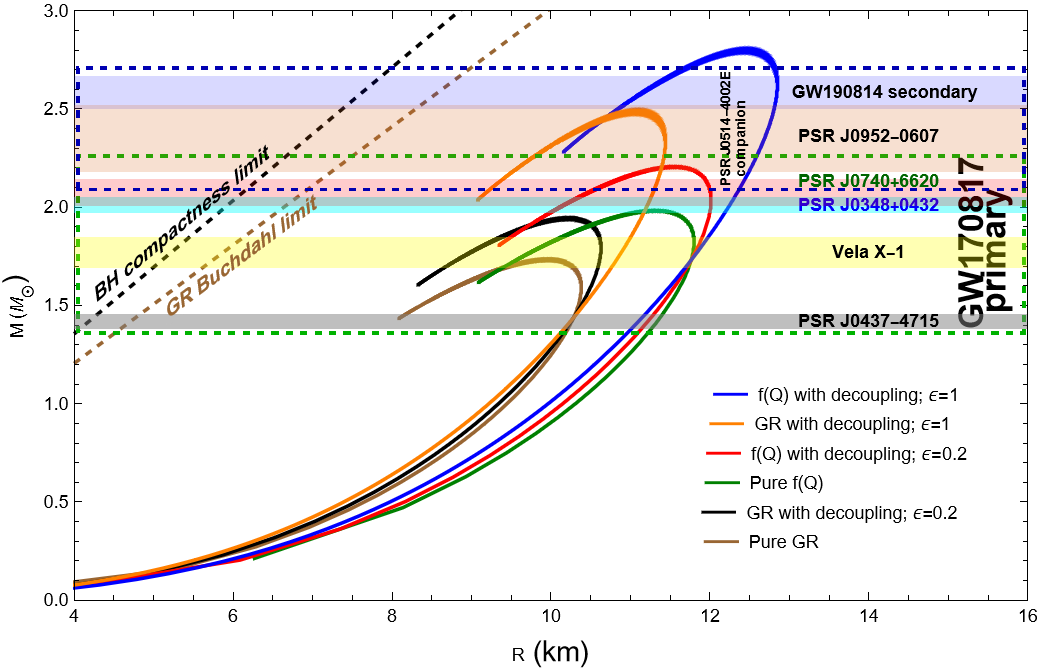}
	\caption{$M–R$ plots in GR and $f(Q)$ gravity (with $\beta_1$=0.8) in presence or absence of the decoupling parameter $\epsilon$ for assumed values of $K=2$ and~$\beta_2=0$ (Set-I).}
	\label{fig15}
\end{figure}

\begin{table}[ht]
\tbl{Values of the model parameters in $GR$ and $f(Q)$ gravity with/without the decoupling parameter $\epsilon$ for the assumed mass and radius of the pulsar PSR~$J0614-3329$ with $K=2$ and $\beta_2=0$.\label{tab3}}
{\begin{tabular}{@{}ccccccc@{}} \toprule
& \multicolumn{3}{c}{GR ($\beta_1=1$)} 
& \multicolumn{3}{c}{$f(Q)$ gravity ($\beta_1=0.8$)} \\
\colrule
$\epsilon$ & $L$ (km) & $C$ & $D$ 
& $L$ (km) & $C$ & $D$ \\
\colrule
$0$   & 23.6154 & 1.91568 & 0.0312205 
      & 23.6154 & 1.91568 & 0.0312205 \\
$0.2$ & 26.6754 & 1.62833 & 0.0312205 
      & 26.6754 & 1.62833 & 0.0312205 \\
$1$ & 36.4299 & 1.34098 & 0.0312205
&36.4299 &  1.34098& 0.0312205\\
\botrule
\end{tabular}}
\end{table}
\begin{table}[ht]
\tbl{EOS parameters for the matter distribution in $GR$ and $f(Q)$ gravity with/without the decoupling parameter $\epsilon$ for the assumed mass and radius of the pulsar PSR~$J0614-3329$ with $K=2$ and $\beta_2=0$.\label{tab4}}
{\begin{tabular}{@{}ccccc@{}} \toprule
& \multicolumn{2}{c}{GR ($\beta_1=1$)} 
& \multicolumn{2}{c}{$f(Q)$ gravity ($\beta_1=0.8$)} \\
\colrule
$\epsilon$ 
& $a$ & $b~(\mathrm{MeV\,fm^{-3}})$ 
& $a$ & $b~(\mathrm{MeV\,fm^{-3}})$ \\
\colrule
$0$   & 0.271845 & 78.1927  & 0.271845 & 62.5542 \\
$0.2$ & 0.355468 & 106.212  & 0.355468 & 84.9700 \\
$1$ & 0.679239& 217.821 & 0.679239  & 174.257 \\
\botrule
\end{tabular}}
\end{table}

In Table~\ref{tab3}, we note that the model parameters remain unchanged when passing from GR to linear $f(Q)$ gravity. This is consistent with the fact that a linear $f(Q)$ gravity does not modify the spacetime geometry; it simply rescales the effective matter sector. In Table~\ref{tab4} and Fig.~\ref{fig12}-\ref{fig14}, we note that the slope of the EOS and the sound speeds remain identical in GR as well as in linear $f(Q)$ gravity models. The only modification introduced by $f(Q)$ gravity is a shift in the surface density. For $\beta_{1} < 1$, the surface density decreases, while for $\beta_{1} > 1$, it increases. Importantly, this shift occurs without altering the EOS slope, sound speed profile or compactness bound. Thus, linear $f(Q)$ induces a uniform vertical displacement of the mass-radius sequence, rather than a structural reshaping of the configuration.

It is important to note that the parameter $\beta_1$ in linear $f(Q)$ gravity enters as an overall scaling factor in the effective energy density and pressure terms (see Eq.~(\ref{rho1})–(\ref{pt1})). Consequently, $\beta_1$ does not alter the geometric structure of the spacetime but rescales the matter sector, leading to a systematic shift in the stellar mass while leaving the geometry unchanged. A detailed discussion of this effect has been presented in Ref.~\cite{Sharma2024}.

A comparative analysis of the mass-radius relationship in GR and in $f(Q)$ gravity is shown in Fig.~\ref{fig15}. The plot illustrates that a linear $f(Q)$ gravity with additional source terms, yields the largest stellar masses for $\beta_{1} < 1$.

\subsection{Compactness bound}
Let us now calculate the maximum compactness bound in our modified gravity theory. To obtain an estimate of the maximum compactness bound, we require that the central pressure not diverge in our model. Using Eq.~(\ref{prf}) and (\ref{ptf}), we note that this condition will be satisfied if we have
\begin{equation}
    LD\sqrt{(K+1)(\epsilon+1)} > C\,(K\epsilon + \epsilon + 1). \label{condition}
\end{equation}
Substituting the values of the constants given in Eq.~(\ref{L})-(\ref{d}), the above constraint leads to the following upper bound on compactness 
\begin{equation}
    u=\frac{M}{R} \leq \frac{2\big[2(K+1)\epsilon + K + 2\big]}{9(K+1)\epsilon + 5K + 9}. \label{bch}
\end{equation}

\begin{figure}[h]
	\centering
	\includegraphics[width=0.6\textwidth]{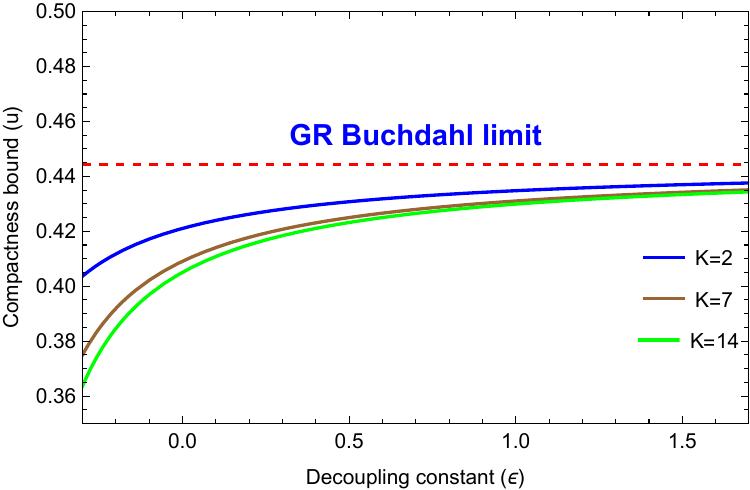}
	\caption{Variation of compactness bound with $\epsilon$ for different values of $K$.}\label{fig16}
\end{figure}

A notable feature of the above bound is that it is independent of the parameters associated with $f(Q)$ gravity. Thus, in a pure linear $f(Q)$ gravity-inspired stellar model, we do not notice any modification in the compactness limit. In the decoupling scenario with an additional source term, however, the compactness limit gets modified through $\epsilon$. In Fig.~\ref{fig16}, we show how the compactness bound varies with $\epsilon$ for different values of $K$. We note that the compactness bound increases monotonically with $\epsilon$ and approaches the Buchdahl limit as $\epsilon$ becomes very large. This behaviour is consistent with our earlier analysis of the Vaidya-Tikekar anisotropic stars~\cite{Chanda2024}, where we observed that the maximum permissible compactness bound decreased with increasing local anisotropy. In the current study, a large value of $\epsilon$ corresponds to a less anisotropic configuration, and hence, a higher compactness for lower anisotropy emerges naturally. Moreover, the upper bound lies well within the Buchdahl bound for the parameter space considered here. In the special case $\epsilon = 0$ (no additional source term) and $K = 0$ (spherical homogeneous matter distribution), the above bound reduces to the well-known Buchdahl bound $u \leq 4/9$.

\subsection{Observational relevance}
One of the main motivations for developing theoretical stellar models is to study compact stars, where relativistic effects cannot be ignored. An important success story in this direction is the development of numerous stellar models capable of describing pulsars, which are broadly clubbed as rotating neutron stars. However, of late, several mass-gap objects (possessing masses greater than the heaviest neutron stars and the lightest black holes) have been detected using multi-messenger observations. Gravitational wave (GW) event $GW190814$ detects an object with mass $\approx2.5-2.67~M_\odot$~\cite{abbott}, while $GW230529$ reports the merger of a neutron Star and a primary mystery compact object with mass in the lower mass-gap region $2.5-4.5~M_\odot$~\cite{sanger}. A recent observation with MeerKAT observatory reports the companion mass of pulsar PSR~$J0514-4002E$ to be $2.09-2.71~M_\odot$~\cite{barr}. 

The current study fits well in the context of the above developments. An intriguing feature of the present model is that for $\epsilon = 1$, the maximum mass reaches $\approx 2.63~M_\odot$ for $\beta_1 = 0.9$ and $\approx 2.8~M_\odot$ for $\beta_1 = 0.8$, placing these configurations within the neutron star-black hole mass-gap region. In contrast, for the same geometric deformation (i.e., the same effective EOS stiffness fixed by $\epsilon$), the maximum mass achievable in GR is $\approx 2.5~M_\odot$. The admissible range of the decoupling parameter extends up to $\epsilon = 1.7$ without violating the causality condition. It is noteworthy that the upper bound on $\epsilon$ is independent of the underlying gravity theory, since for fixed $\epsilon$ the sound speeds and hence the EOS stiffness remain identical in GR and linear $f(Q)$ gravity. Although the model formally allows stiffness approaching the causal bound as $\epsilon$ increases, realistic neutron star equations of state inferred from nuclear physics and astrophysical observations typically do not. Therefore, within the physically realistic stiffness regime (i.e., well below the causal limit), linear $f(Q)$ gravity still supports systematically heavier stellar configurations than GR under identical geometric deformation and EOS stiffness.

Let us now clarify the distinct roles played by the parameters $\beta_{1}$ and $\epsilon$. In our linear $f(Q)$ gravity model, the field equations themselves are rescaled. Consequently, the energy density, two pressures, anisotropy, surface density, etc., decrease when the value of  $\beta_1$ is decreased, as can be seen in Fig.~\ref{fig8}-\ref{fig11} and Table~\ref{tab4}. This makes it possible for a pure $f(Q)$ gravity model with $\beta_{1} < 1$ to support larger masses than in GR and even to reach the neutron star-black hole mass gap, without altering the EOS stiffness or the underlying VT$+$Karmarkar geometry. On the other hand, since the metric potentials remain fixed in a linear $f(Q)$ theory, the maximum compactness bound and the geometric structure of the equilibrium configuration do not change. 

The gravitational decoupling parameter $\epsilon$ provides an additional degree of freedom that is absent in pure $f(Q)$ gravity. It introduces an explicit deformation of the metric potentials, thereby modifying both the geometry and matter simultaneously. The geometric deformation alters the anisotropic profile, the hydrostatic balance and the admissible compactness, enabling the construction of configurations with significantly higher masses than those allowed by varying $\beta_{1}$ alone. As a result, for a fixed $\beta_1$, increasing $\epsilon$ continues to increase the maximum mass beyond what pure linear $f(Q)$ can achieve. Thus, the decoupling approach extends the accessible mass range and allows for even heavier, more compact stars within the same underlying gravitational theory. Therefore, while a linear $f(Q)$ gravity can rescale the total gravitational mass through the parameter $\beta_{1}$, the geometric deformation induced by $\epsilon$ can provide higher mass objects. The model, thus, can accommodate a much wider class of compact stars than in GR. The enhancement of the mass window arises from two independent mechanisms: coupling-driven matter rescaling ($\beta_1$) and geometric deformation ($\epsilon$). Their independent action allows configurations unattainable in GR for the same effective stiffness. 

In Table~$5$, we show that for fixed $\beta_1 (=0.9)$, how the decoupling parameter $\epsilon$ can be fine-tuned to make the model compatible with observed data of some of the well-known pulsars. When there is uncertainty in the radius measurement, if the pulsar's mass is well estimated, this technique can also be used to estimate the corresponding radius, as shown in the table.

Dependence of the predicted stellar radius on the decoupling parameter $\epsilon$ for the observed compact objects, including some well-measured pulsars 
PSR~$J0740+6620$ ($M = 2.08 \pm 0.07\,M_\odot$, $R = 13.7^{+2.6}_{-1.5}\,\mathrm{km}$~\cite{miller};
$M = 2.073 \pm 0.069\,M_\odot$, $R = 12.49^{+1.28}_{-0.88}\,\mathrm{km}$~\cite{salmi}), 
PSR~$J0348+0432$ ($M = 2.01 \pm 0.04\,M_\odot$~\cite{anto};
$R = 12.246$-$12.957\,\mathrm{km}$, model dependent~\cite{zhao2016}), 
PSR~$J0437-4715$ ($M = 1.418 \pm 0.037\,M_\odot$, $R = 11.36^{+0.95}_{-0.63}\,\mathrm{km}$~\cite{choudhury2024}), 
and PSR~$J0952-0607$ ($M = 2.35 \pm 0.17\,M_\odot$~\cite{romani};
$R = 13.21 \pm 0.96\,\mathrm{km}$~\cite{elhanafy2024}) is summarized in Table~\ref{tab:pulsar_eps_compare}.

These results indicate that larger values of $\epsilon$ allow the construction of more massive stellar configurations. In particular, pulsars with masses above $2~M_\odot$ require $\epsilon \gtrsim 0.2$ for the specific parameter set used in this analysis. For different values of $\beta_1$, the required decoupling parameter to match the observed mass-radius data shifts accordingly. Nevertheless, the qualitative trend remains consistent i.e, stronger coupling supports heavier stars. Table~\ref{tab:pulsar_eps_compare} offers an exploratory assessment of how varying $\epsilon$ tunes the predicted radii while keeping $K$, $\beta_1$, and $\beta_2$ fixed.

A comparative data set in GR and $f(Q)$ gravity is presented in Table~\ref{tab:pulsar_eps_compare}. We note that in the presence of the additional source, a linear $f(Q)$ gravity with $\beta_1 < 1$ predicts systematically larger mass and radii than in GR. For relatively lighter pulsars such as PSR~$J0437-4715$, the observed mass band is better reproduced within pure linear $f(Q)$ gravity ($\beta_1=0.8$) without requiring strong geometric deformation. In contrast, heavier pulsars like PSR $J0740-6620$ require the combined effect of matter-sector rescaling and significant geometric deformation ($\epsilon$=1). For very massive objects such as PSR~$J0952-0607$ and candidates in the neutron star-black hole mass-gap region, GR fails to accommodate the observed mass range within our chosen parameter domain, even under stronger decoupling. These objects, however, are supported in linear $f(Q)$ gravity with the same EOS stiffness (fixed $\epsilon$) as in GR, owing to the independent matter-sector rescaling governed by $\beta_1$. This demonstrates that linear $f(Q)$ gravity does not merely replicate GR solutions but enlarges the admissible stellar mass window through a controlled coupling-driven shift, while gravitational decoupling independently regulates the geometric deformation and effective stiffness.

Our investigation demonstrates that linear $f(Q)$ gravity produces a controlled coupling-driven shift of the mass-radius sequence at fixed geometry, while gravitational decoupling independently modifies geometric compactness. The two effects are structurally distinct and cannot be reduced to a single deformation parameter.

\begin{table}[ht]
\tbl{Predicted stellar radii for observed compact objects with varying values of $\epsilon$ with model parameters $\beta_1=0.9$, $K=2$, $\beta_2=0$.\label{6tab:pulsar_eps}}
{\begin{tabular}{lccccc} \toprule
Compact object & $M_{\mathrm{obs}}/M_{\odot}$ & $R_{\mathrm{obs}}$ (km) & $\epsilon$ & $\approx R_{\mathrm{pred}}$ (km) \\
\colrule

PSR~J0740+6620
& $2.073\pm0.069$~\cite{salmi}
& $12.49^{+1.28}_{-0.88}$~\cite{miller,salmi}
& $0.5$ & $11.55$-$11.61$ \\
& & & $1$ & $11.69$-$11.85$ \\
\colrule

PSR~J0348+0432
& $2.01\pm0.04$~\cite{anto}
& -
& $0.2$ & $11.01$-$11.23$ \\
& & & $0.5$ & $11.52$-$11.58$ \\
& & & $1$ & $11.65$-$11.76$ \\
\colrule

PSR~J0437$-$4715
& $1.418\pm0.037$~\cite{choudhury2024}
& $11.36^{+0.95}_{-0.63}$~\cite{choudhury2024}
& $-0.2$ & $11.01$-$11.06$ \\
& & & $0$ & $10.70$-$10.81$ \\
& & & $0.1$ & $10.67$-$10.79$ \\
& & & $0.2$ & $10.61$-$10.77$ \\
& & & $0.5$ & $10.60$-$10.75$ \\
& & & $1$ & $10.52$-$10.70$ \\
\colrule

Vela $X-1$
& $1.77\pm0.08$~\cite{Rawls2011}
& -
&  $0.1$ & $11.06$-$11.11$ \\
& & & $0.2$ & $11.11$-$11.25$ \\
& & & $0.5$ & $11.14$-$11.39$ \\
& & & $1$ & $11.19$-$11.45$ \\
\colrule

PSR~J0952$-$0607
& $2.35\pm0.17$~\cite{romani}
& -
& $1$ & $11.90$-$12.07$ \\
\colrule
GW170817 primary
& $1.36-2.26\,M_\odot$~\cite{Abbott2017}
& -
& $0.5$ & $10.53$-$10.60$ \\
&&& $1$ & $10.47$-$11.98$ \\
\colrule

GW190814 mass gap object
& $2.59^{+0.08}_{-0.09}\,M_\odot$~\cite{abbott}
& -
& $1$ & $11.35$-$11.73$ \\
\colrule
PSR~$J0514-4002E$ companion
&$2.09-2.71$~\cite{barr}
& -
& $0.5$ & $\ge11.6$\\
&&&  $1$ & $\ge 11.8$\\

\botrule
\end{tabular}}
\end{table}

\begin{sidewaystable}[p]
\begin{center}

\caption{Comparison of predicted stellar radii in GR and $f(Q)$ gravity for selected compact objects with/without decoupling parameter for $K=2$, $\beta_2=0$.}
\label{tab:pulsar_eps_compare}

\setlength{\tabcolsep}{6pt}
\renewcommand{\arraystretch}{1.2}

\begin{tabular}{p{3cm} p{2.2cm} p{2cm} c c c c}
\hline
Compact object & $M_{\mathrm{obs}}/M_{\odot}$ & $R_{\mathrm{obs}}$ (km) & $\epsilon$
& $R_{\mathrm{GR}}~(km)$ 
& $R_{f(Q),0.9}~(km)$ 
& $R_{f(Q),0.8}~(km)$ \\
\hline

PSR J0437-4715
& $1.418\pm0.037$ 
& $11.36^{+0.95}_{-0.63}$
& 0   
& $10.22$-$10.30$ 
& $10.70$-$10.81$ 
& $11.18$-$11.40$ \\

&  
&  
& 0.2 
& $10.20$-$10.32$ 
& $10.61$-$10.77$ 
& $11.08$-$11.30$ \\

&  
&
& 1 
& $10.16$-$10.30$ 
& $10.52–10.70$ 
& $11.01$-$11.18$ \\
\hline

PSR J0740-6620
& $2.073\pm0.069$
& $12.49^{+1.28}_{-0.88}$
& 0   
& Not in range  
& Not in range 
& Not in range \\
&
&
& 1 
& $11.21-11.36$ 
& $11.69–-11.85$
& $12.25-12.45$ \\
\hline

PSR~J0348+0432
& $2.01\pm0.04$
& -
& 0   
& Not in range  
& Not in range 
& Not in range \\
&
&
& 1 
& $11.16-11.25$
& $11.65–-11.76$
&$12.19-12.30$\\
\hline

PSR~J0952$-$0607
& $2.35\pm0.17$
& -
& 0   
& Not in range  
& Not in range 
& Not in range \\
&
&
& 1 
& Not in range  
& $\ge 11.36$ 
& $12.48-12.82$ \\
\hline

GW190814 mass-gap object
& $2.59^{+0.08}_{-0.09}$
& -
& 0   
& Not in range  
& Not in range 
& Not in range \\
&
&
& 1 
& Not in range  
& Not in range 
& $12.36-12.84$ \\
\hline

PSR~$J0514-4002E$ companion
&$2.09-2.71$
& -
& 0   
& Not in range  
& Not in range 
& Not in range \\
&
&
& 1 
& $\ge11.32$  
& $\ge11.8$ 
& $12.36-12.83$ \\
\hline

\end{tabular}
\end{center}
\end{sidewaystable}

\section{Concluding Remarks}\label{sec6}
The present analysis clarifies that linear $f(Q)$ gravity alone does not generate new geometric families of compact stars, as its metric sector is dynamically equivalent to GR. However, when combined with gravitational decoupling, the theory acquires a genuine two-parameter deformation structure. The parameter $\beta_1$ uniformly rescales matter variables, while $\epsilon$ induces geometric deformation that modifies both geometry and matter, reduces anisotropy, stiffens the effective equation of state (EOS) and the maximum permissible compactness. This geometric deformation elevates the entire mass-radius sequence and opens access to $> 2~M_{\odot}$ stars, which is particularly interesting in the context of recently observed stellar masses in the mass-gap region. A recent analysis of Alwan \textit{et al}~\cite{alwan2025} shows that rotational observables such as the moment of inertia and the $\bar{I}\!-\!C$ quasi-universal relation are highly sensitive to changes in the stellar interior in $f(Q)$ gravity. This suggests that the interior modifications generated here through the decoupling parameter $\epsilon$ may likewise influence strong-field rotational signatures used to test $f(Q)$ gravity. Such studies could point towards observational signatures that could help discriminate between different gravitational theories and interior structures. The prospect of such observational signatures is promising in the light of current multi-messenger astronomy. For example, the MeerKAT observatory pulsar timing data have recently revealed unusually massive neutron stars whose properties challenge our current understanding of physics~\cite{barr}. Several GW events also point towards objects in the neutron star-black hole mass-gap region. It is interesting to note that the current model can accommodate a wide range of stellar masses: a modest deformation reproduces the behaviour of well-measured millisecond pulsars. A stronger coupling naturally supports the heavier compact objects highlighted by radio timing and gravitational wave observations. Predicted radii for sources like PSR~$J0740+6620$, PSR~$J0348+0432$, PSR~$J0437-4715$.\\
To conclude, we show that a combination of linear $f(Q)$ gravity, embedding class-I geometry, and gravitational decoupling offers a flexible, physically transparent, and observationally compatible description of compact stars across the full spectrum of observed pulsars, from ordinary neutron stars to emerging mass-gap candidates. The framework, therefore, extends embedding class-I VT models from a single-parameter GR deformation scheme to a structurally richer two-parameter system, enabling controlled investigation of coupling-driven mass enhancement under identical geometric conditions. The novelty of the current investigation are summarized below:

\medskip
\begin{itemize}
	
	\item Previous embedding class-I Vaidya-Tikekar (VT) constructions in linear $f(Q)$ gravity (e.g., Ghosh \textit{et al}, 2024) are geometrically equivalent to GR and differ only through uniform matter-sector rescaling by $\beta_1$. The present work introduces gravitational decoupling within this framework, thereby enlarging the solution space beyond pure normalization effects and generating a genuine geometric deformation sector.
	
	\item In contrast to GR-based VT + decoupling models, which contain only a single deformation parameter, the present framework possesses two independent parameters $(\epsilon,~\beta_1)$. The parameter $\epsilon$ deforms the metric and modifies effective EOS stiffness, while $\beta_1$ independently rescales the matter sector without altering the geometry.
	
	\item This separation enables a controlled structural decomposition of mass enhancement mechanisms: geometric stiffening driven by $\epsilon$ versus coupling-driven matter rescaling governed by $\beta_1$.
	
	\item A direct comparison between GR and linear $f(Q)$ gravity is performed at identical geometric deformation (fixed $\epsilon$), thereby isolating the sole effect of matter-sector rescaling.
	
	\item It is demonstrated that, at fixed EOS stiffness, linear $f(Q)$ systematically shifts the mass-radius sequence relative to GR, yielding larger maximum masses through coupling-induced surface density modification.
	
	\item An analytic compactness bound is derived for the decoupled embedding class-I configuration, showing explicitly that compactness is modified by geometric deformation ($\epsilon$) but remains independent of the linear $f(Q)$ coupling parameter.
	
	\item The combined action of $(\epsilon,~\beta_1)$ enlarges the admissible stellar mass window within causality and regularity limits, allowing configurations compatible with high-mass pulsars and mass-gap candidates without invoking super-causal stiffness.
	
\end{itemize}

\section*{Acknowledgments}
We express our sincerest thanks to the anonymous referee for constructive suggestions. RS gratefully acknowledges support from the Inter-University Centre for Astronomy and Astrophysics (IUCAA), Pune, India, under its Visiting Research Associateship Programme.

\end{document}